\documentclass[floatfix,aps,prd,twocolumn,letterpaper,lengthcheck,superscriptaddress,showpacs,amssymb,amsmath,amsfonts,nofootinbib,altaffilletter,nopreprintnumbers,showpacs,longbibliography]{revtex4-1}
\usepackage[utf8]{inputenc}
\usepackage{xspace}
\usepackage{acronym}
\usepackage{booktabs}
\usepackage{graphicx}
\usepackage{amssymb}
\usepackage{color}
\usepackage{xcolor}
\usepackage{wasysym}
\usepackage{url}
\usepackage[colorlinks, linkcolor=blue, pdfborder={0 0 0}, breaklinks=true]{hyperref}
\definecolor{lightgray}{gray}{0.9}
\usepackage{tabularx}
\usepackage{multirow}
\usepackage{scrextend}
\usepackage{textgreek}
\usepackage{graphicx}
\usepackage[caption=false]{subfig}
\usepackage{mathtools}
\usepackage{float}
\usepackage{soul}
\usepackage{booktabs}
\usepackage{enumitem}
\usepackage{dcolumn}
\usepackage{bm}
\usepackage{lineno}

\newcommand{\qm}[1]{\color{black} #1}

\DeclarePairedDelimiterX{\norm}[1]{\lVert}{\rVert}{#1}

\begin{document}

\preprint{APS/123-QED}

\title{Gravitational-Wave Searches for Cosmic String Cusps in Einstein Telescope Data using Deep Learning}

\author{Quirijn Meijer}\email[Corresponding author: ]{r.h.a.j.meijer@uu.nl}
\affiliation{Institute for Gravitational and Subatomic Physics (GRASP), Department of Physics, Utrecht University,
Princetonplein 1, 3584CC Utrecht, The Netherlands}
\affiliation{Nikhef, Science Park 105, 1098XG Amsterdam, The Netherlands}

\author{Melissa Lopez}
\affiliation{Institute for Gravitational and Subatomic Physics (GRASP), Department of Physics, Utrecht University,
Princetonplein 1, 3584CC Utrecht, The Netherlands}
\affiliation{Nikhef, Science Park 105, 1098XG Amsterdam, The Netherlands}

\author{Daichi Tsuna}
\affiliation{TAPIR, Mailcode 350-17, California Institute of Technology, Pasadena, California 91125, USA}
\affiliation{Research Center for the Early Universe (RESCEU), School of Science, The University of Tokyo, 7-3-1 Hongo, Bunkyo-ku, Tokyo 113-0033, Japan}

\author{Sarah Caudill}
\affiliation{Department of Physics, University of Massachusetts, Dartmouth, Massachusetts 02747, USA}
\affiliation{Center for Scientific Computing and Data Science Research, University of Massachusetts, Dartmouth, Massachusetts 02747, USA}

\date{\today}

\begin{abstract}
\noindent Gravitational-wave searches for cosmic strings are currently hindered by the presence of detector glitches, some classes of which strongly resemble cosmic string signals. This confusion greatly reduces the efficiency of searches. A deep-learning model is proposed for the task of distinguishing between gravitational-wave signals from cosmic string cusps and simulated blip glitches in design sensitivity data from the future Einstein Telescope. The model is an ensemble consisting of three convolutional neural networks, achieving an accuracy of $79\%$, a true positive rate of $76\%$, and a false positive rate of $18\%$. This marks the first time convolutional neural networks have been trained on a realistic population of Einstein Telescope glitches. On a dataset consisting of signals and glitches, the model is shown to outperform matched filtering, specifically being better at rejecting glitches. The behaviour of the model is interpreted through the application of several methods, including a novel technique called waveform surgery, used to quantify the importance of waveform sections to a classification model. In addition, a method to visualise convolutional neural network activations for one-dimensional time series is proposed and used. These analyses help further the understanding of the morphological differences between cosmic string cusp signals and blip glitches. Because of its classification speed in the order of magnitude of milliseconds, the deep-learning model is suitable for future use as part of a real-time detection pipeline. The deep-learning model is transverse and can therefore potentially be applied to other transient searches.
\end{abstract}

\maketitle

\section{Introduction}

Since the first confirmed detection of the gravitational-wave signal GW150914 in 2015 \cite{GW150914}, over $90$ gravitational waves have been confirmed by the LIGO, Virgo and KAGRA detectors \cite{PhysRevX.9.031040, PhysRevX.11.021053, theligoscientificcollaboration2021gwtc3, gwtc21}. These observatories are currently in their second generation \cite{AdvancedLIGO, AdvancedVIRGO}. A third generation of detectors including Cosmic Explorer \cite{CE}, the Laser Interferometer Space Antenna (LISA) \cite{LISA} and the Einstein Telescope \cite{ET} are already in development. The Einstein Telescope will have a greatly increased sensitivity compared to the current generation and is expected to detect many more signals, possibly from new sources. Gravitational waves observed thus far have been the product of compact binary coalescences, which are pairs of coalescing stellar- or intermediate-mass black holes and neutron stars \cite{PhysRevX.9.031040, PhysRevX.11.021053, theligoscientificcollaboration2021gwtc3, gwtc21}. Searches, however, are not limited to such systems. One class of unary sources is that of cosmic strings.

Cosmic strings are objects that are conjectured by several theories to have formed in the early Universe, and if present, have evolved as the Universe expanded \cite{Vilenkin, Kibble:1976sj}. They should present themselves as strings at cosmological scales. Cosmic strings interact with gravity through gravitational lensing on background light sources due to their angular deficit \cite{Vilenkin}, but also through gravitational waves. The focus of this paper is the detection of cusps on cosmic strings \cite{2005PhRvD..71f3510D, 2000PhRvL..85.3761D, 2001PhRvD..64f4008D}. Cusps can be understood as points on the cosmic string that {\qm instantaneously} accelerate to the speed of light, and in doing so generate gravitational waves.

Current searches for cosmic string signatures, of which cusp signals are an example, rely on matched filtering \cite{LVKO3CS, Allskybursts, Siemens, Aasi_2014, Abbott_2009, Abbott_2018}. Matched filtering is a process where modelled waveforms (called templates) are convolved with detector strain data in order to check for the presence of a signal matching the template. Although these searches have not resulted in observational evidence for the existence of cosmic strings, their results have been used to constrain the model parameters of cosmic strings \cite{Aasi_2014, LVKO3CS, Abbott_2018}. Matched filter searches for cosmic strings are hindered by the presence of detector glitches \cite{Allskybursts, Abbott_2018}, bursts of non-Gaussian noise that may look very similar to modelled cusp signals. Although it is uncertain what glitches will look like in the Einstein Telescope, short-duration glitches that mimic cusp signals are likely to appear. In this paper, machine learning is employed to demonstrate it is possible to differentiate cosmic string cusps from a common class of transient glitches known as blip glitches in LIGO and Virgo data \cite{Cabero_2019}, assuming similar glitches in Einstein Telescope data.

This paper details the training of convolutional neural networks for the task of distinguishing modeled cosmic string cusp signals from artificial blip glitches in simulated Einstein Telescope data. The goal is to both prepare for the arrival of the third generation of detectors, as well as to utilise the higher sensitivity of these detectors to learn about the morphological differences between the two types when obfuscated by detector noise. Having this information may aid in current searches in second-generation data as well, as it can be incorporated to design better searches and confirmation tests for observed gravitational-wave candidates.

This paper is organised as follows. Section \ref{sec:cosmicstrings} reviews cosmic strings before drawing the comparison to glitches through their waveform similarity. Section \ref{sec:currentmethods} reviews matched filtering, the current method for cosmic string searches. Section \ref{sec:methodology} details the methodology of this paper, from the creation of the dataset to the analysis of the model. Section \ref{sec:results} reports on the results of the applied methods. Conclusions are collected in Sec. \ref{sec:conclusions}.

\section{Cosmic Strings and Glitches}\label{sec:cosmicstrings}

\begin{figure}[!]
    \includegraphics[width=.3\columnwidth]{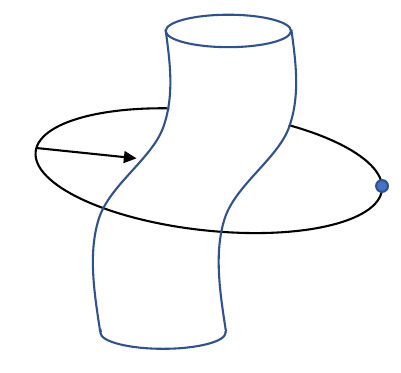}%
    \caption{A contraction of the circle $S^{1}$ to the basepoint on the right, caught on a stringlike deficiency. This deficiency, or defect, does not allow the contraction to complete.}
    \label{fig:contraction}
\end{figure}

Cosmic strings are found in field theories, where they appear as one-dimensional topological defects \cite{Vilenkin, Kibble:1976sj}. Such defects may arise as the result of a process called spontaneous symmetry breaking, where the internal symmetry group of the vacuum manifold $M$ is lowered to a strict subgroup \cite{Folland}. {\qm Although both global and local symmetries can be broken the restriction to local symmetry breaking is made, due to the possible relation with unification} \cite{Sakellariadou}. It is for this reason that cosmic strings originating from symmetry breaking in local symmetry groups, or gauge groups \cite{peskin1995introduction}, are studied in this paper. Assuming the presence of a Lie group structure leads to the gauge group being a manifold, and in particular to the gauge group admitting a topology. It is through the homotopy groups \cite{Bredon} of the gauge group that topological defects can be detected and classified. In particular, the fundamental group $\pi_{1}(M)$ being non-trivial leads to the conclusion that effectively one-dimensional (or stringlike) topological defects must be present in the theory, as the contractions of the $S^{1}$ embeddings get caught on such presences (illustrated in Fig. \ref{fig:contraction}). As the circle shrinks, the defect prohibits the circle from collapsing onto the base point. Different defects are then signaled by the classes in the fundamental group. More generally, non-triviality of the $k$-th homotopy group demonstrates the presence of topological defects of dimension $k$. Although cosmic strings remain hypothetical as of yet, the detection of topological defects in other dimensions {\qm for other systems} gives reason to assume they may exist. Domain walls, two-dimensional topological defects, appear when a ferromagnetic material undergoes a phase transition as its temperature passes the Curie point \cite{Schwarz}. 

\begin{figure}[!]
    \includegraphics[width=.7\columnwidth]{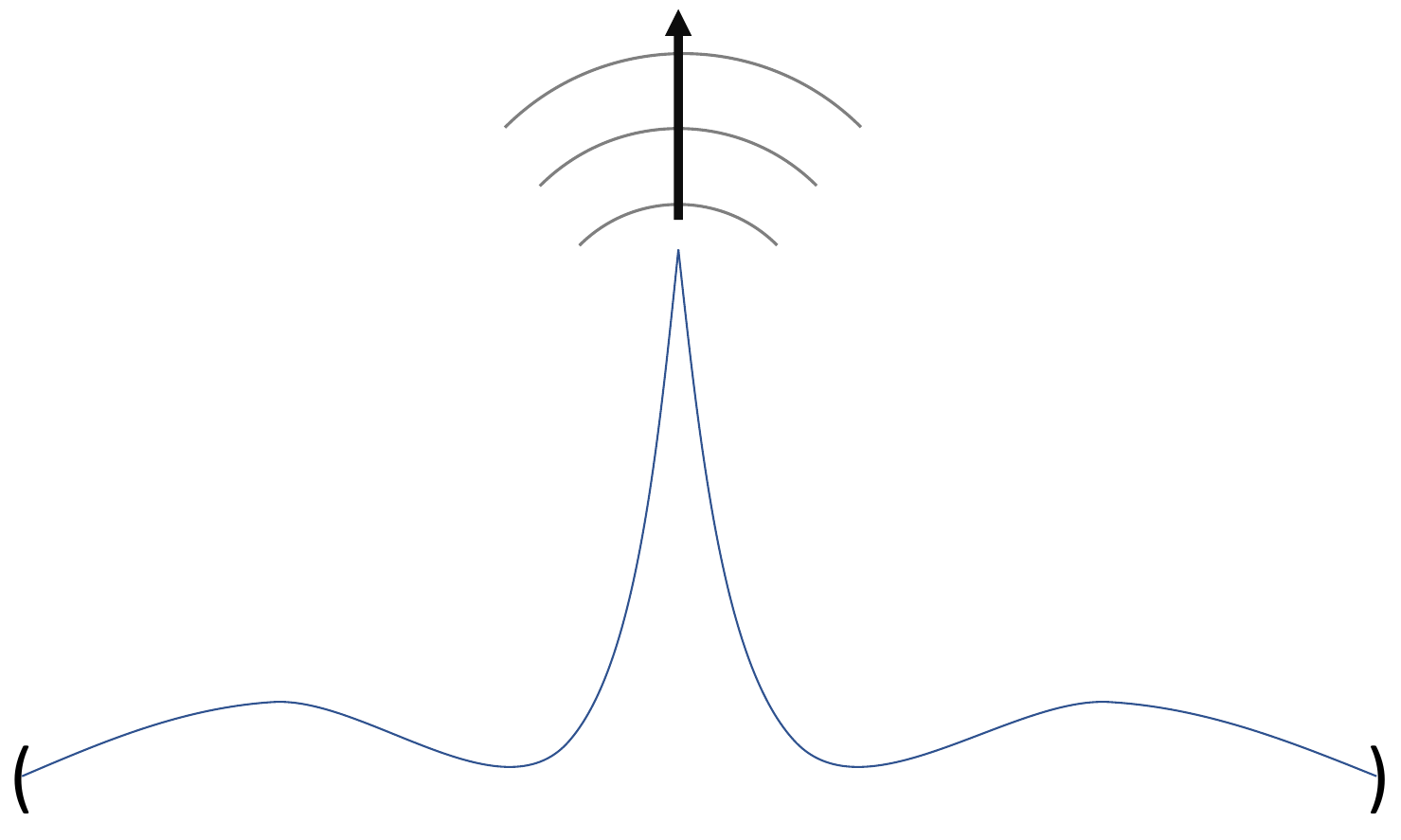}%
    \caption{A visualisation of a burst gravitational wave produced at a cosmic string cusp. As the string snaps into a cusp, a directed gravitational wave is emitted in the direction of acceleration.}
    \label{fig:emission}
\end{figure}

Alternatively, a class of cosmic strings arises from string theory. In string theory, strings are small elemental objects that vibrate in dimensions beyond the four spacetime dimensions postulated by general relativity \cite{Polchinski}. As these additional dimensions are compactified (for instance through the Kaluza-Klein mechanism \cite{DUFF19861}), this takes place at unobservably small scales, meaning it is extremely difficult to obtain observational evidence. However, it is possible for these strings to grow to a cosmological scale, forming so-called cosmic superstrings that exhibit behaviour similar to cosmic strings \cite{Copeland, Sakellariadou}.

Spontaneous symmetry breaking \cite{peskin1995introduction}, and therefore the appearance of cosmic strings, may be caused by phase transitions such as the ones associated with grand unification or lower-energy scales. Cosmic strings are therefore of interest to the scientific community as their study can unveil information about both the early Universe and a string-theoretical description of the Universe \cite{Sakellariadou}.

As physical phenomena, cosmic strings appear at cosmological scale as extremely thin strings with massive densities. As such, their large-scale dynamics are governed by the zero-thickness limit by the Nambu-Goto action \cite{Sakellariadou}. Cosmic strings can either be open strings or closed loops and moreover may interact if two cosmic strings meet. Networks of multiple interacting cosmic strings have been simulated {\qm \cite{PhysRevD.89.023512, LarissaLorenz_2010, PhysRevD.40.973, PhysRevD.41.2408, PhysRevLett.64.119}}. 

In order to detect cosmic strings, observational signatures are needed. Cosmic strings are massive dynamic objects, producing gravitational waves through a variety of mechanisms. Examples are the formation of cusps and kinks \cite{LVKO3CS}. This work focuses on cusps in closed cosmic strings. A cusp is a singular point on a curve where the tangent vector vanishes, or in other words, a singularity where a point traveling along the curve would have to reverse its direction. When this happens, the physical string snaps into a cusp shape and is at that point {\qm instantaneously} accelerated to the speed of light. A burst gravitational wave is then emitted in the direction of acceleration \cite{2005PhRvD..71f3510D, 2000PhRvL..85.3761D, 2001PhRvD..64f4008D}. This is visualised in Fig. \ref{fig:emission}. The waveform $h$ of such a {\qm signal in the Nambu-Goto limit} for loop length $l$ at redshift $z$ and tension $G \mu$, {\qm in natural units where the speed of light $c$ is taken to be unity}, {\qm has been computed as a function of frequency $f$ as \cite{LVKO3CS}:}

\begin{equation}
    \begin{split}
        h_{l, z, G\mu}(f) & = \left[ (2/3)^{2/3} 8/\Gamma^{2}(1/3) \frac{l^{2/3} G\mu}{(1+z)^{1/3} r(z)} \right] f^{-4/3} \\
        & \approx \left[ 0.85 \frac{l^{2/3} G\mu}{(1+z)^{1/3} r(z)} \right] f^{-4/3}.
    \end{split}
    \label{eq:cuspwaveform}
\end{equation}

\begin{figure}[!]
    \captionsetup[subfigure]{labelformat=empty}
    \subfloat[\label{fig:compsignalstrain}]{%
    \includegraphics[width=0.5\columnwidth]{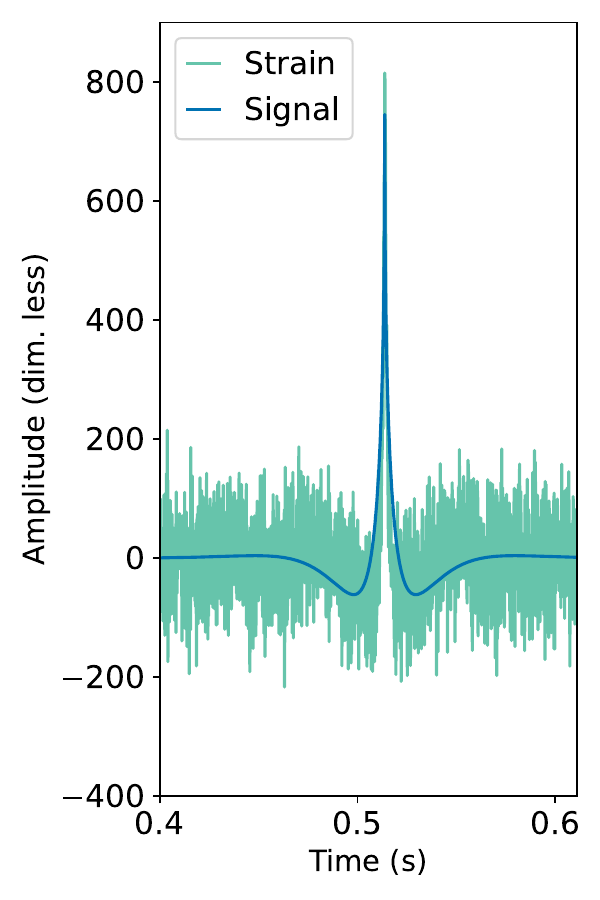}%
}\hfill
    \subfloat[\label{fig:compsignalspect}]{%
    \includegraphics[width=0.5\columnwidth]{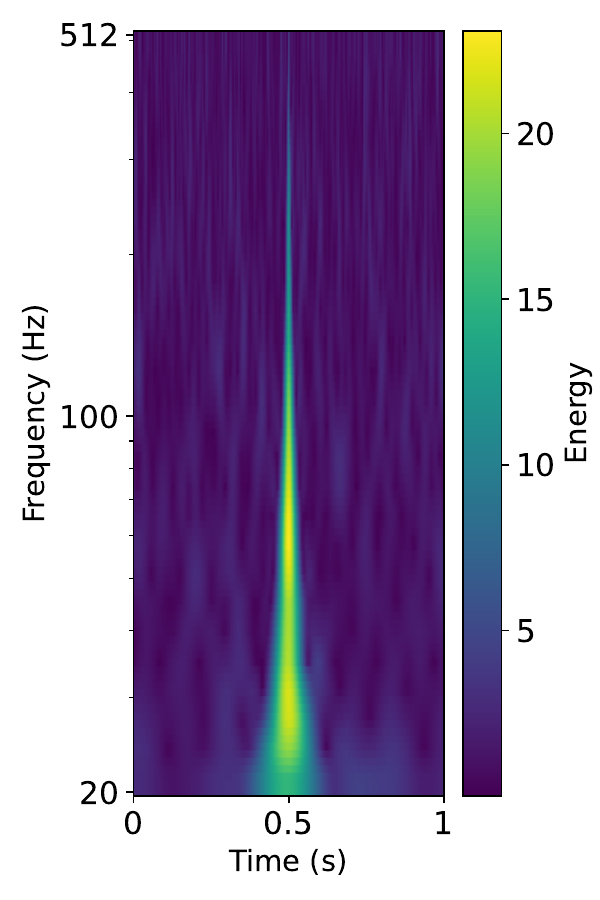}%
}\hfill
    \caption{{\qm A cusp signal with an amplitude of approximately $9.85 \times 10^{-22} \;\textup{Hz}^{1/3}$ prior to injection, overlaid onto the noise it was injected into on the left, and the spectrogram of this strain on the right. Note that the amplitude absorbs the parameters $l$, $z$ and $G\mu$ per Eq. \ref{amplitudepre}.}}
    \label{fig:compsignal}
\end{figure}

\noindent In this formula $r(z)$ is the comoving distance to the loop, or the distance of the observer to the loop, and $\Gamma$ is the Gamma function. The extrinsic parameters for detection are distance and the sky location. The shape of this waveform in the time domain, and a spectrogram of a strain {\qm of noise} including this waveform, are shown in Fig. \ref{fig:compsignal}.

\begin{figure}[!]
    \captionsetup[subfigure]{labelformat=empty}
    \subfloat[\label{fig:compglitchstrain}]{%
    \includegraphics[width=0.5\columnwidth]{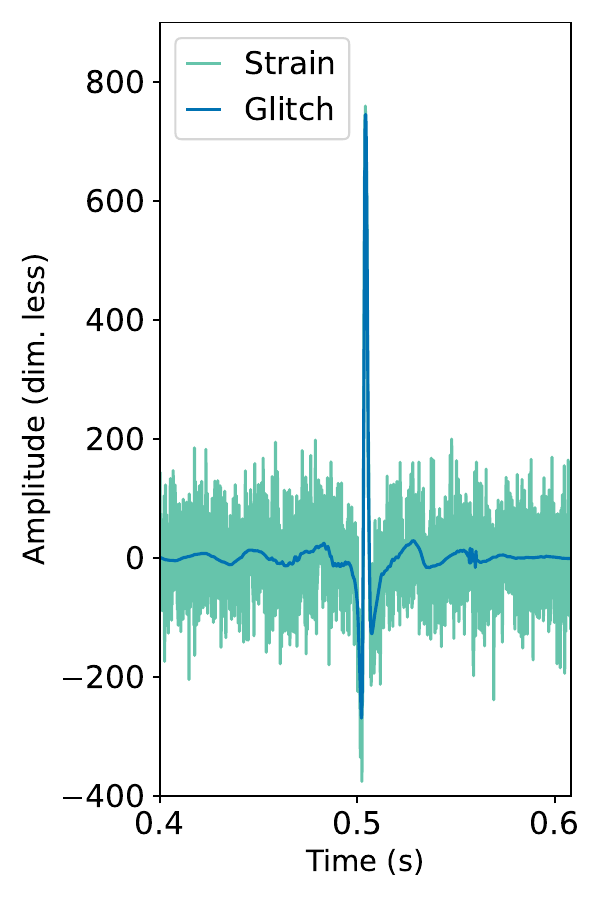}%
}\hfill
    \subfloat[\label{fig:compglitchspect}]{%
    \includegraphics[width=0.5\columnwidth]{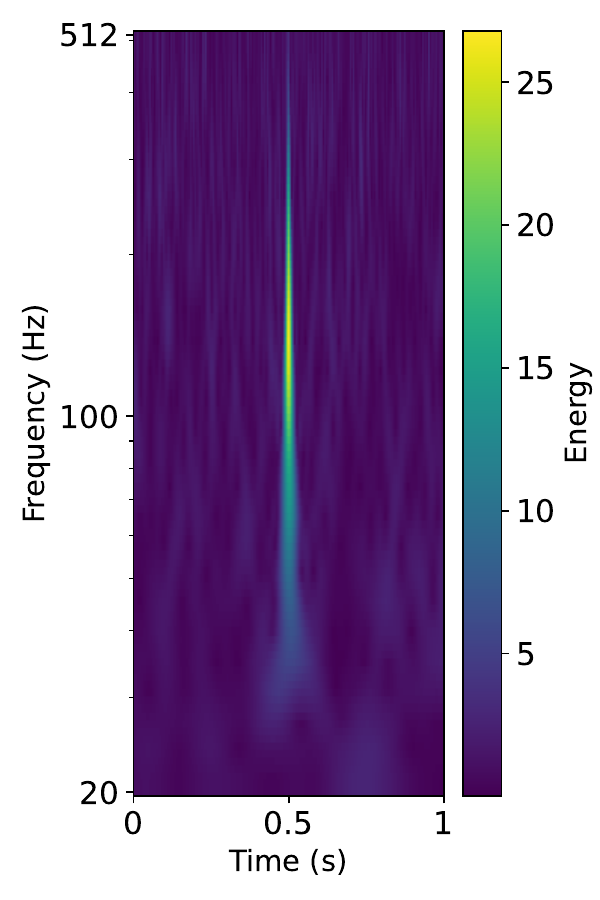}%
}\hfill
    \caption{{\qm A glitch generated by \texttt{gengli}, overlaid onto the noise it was injected into on the left, and the spectrogram of this strain on the right. The glitches were scaled to follow the SNR distribution of the cusp signals. This procedure is explained in Sec. \ref{sec:dataset}.}}
    \label{fig:compglitch}
\end{figure}

State-of-the-art methods employed in cosmic string searches such as matched filtering (reviewed in Sec. \ref{sec:currentmethods}) are hindered by the similarity of cosmic string cusp signals to short-duration transient glitches like blip glitches. Blip glitches are defined as transient bursts with a duration of around $25$ ms with frequency concentrated between $30$ and $250$ Hz \cite{Cabero_2019}. Depending on the viewing angle and assumptions on loop length \cite{Abbott_2018}, a cosmic string cusp signal may occupy this same frequency range. This paper is focused on the development of methods with respect to blip glitches. However, the methods treated could be extended to any class of short-duration glitches affecting cosmic string cusp searches. Although the morphology of such glitches can differ strongly from cusp signals, in the worst-case scenario they may look near-identical, especially when accounting for the diffusion caused by background noise. This worst-case likeness is demonstrated in Figs.~\ref{fig:compsignal} and~\ref{fig:compglitch}.

\section{Current Methods}\label{sec:currentmethods}

Current state-of-the-art methods rely on matched filtering, which is optimal for finding known signals in the presence of stationary and Gaussian noise \cite{helstrom1960statistical}. Matched filtering convolves a known signal (or filter) with a data segment in order to obtain a signal-to-noise ratio (SNR) value that indicates the presence of the signal in the data. If the value so obtained exceeds a preset threshold, it is said the filter was matched to the data, and the GPS time of the trigger is stored. In gravitational-wave pipelines, this trigger is the starting point for a series of statistical tests to confirm a gravitational-wave candidate {\qm \cite{LVKO3CS, 2021SoftX..1400680C}}.

Given a linear space of complex functions {\qm into which the waveforms can be embedded}, {\qm the SNR is dependent on the following Hermitian inner product for $u$ and $v$ taken from this space:}

\begin{equation}
    \langle u, v \rangle = 4 \textup{Re}\left[ \int_{0}^{\infty} \frac{u(f) \overline{v(f)}}{\textup{S}_{n}(f)} \mathrm{d}f \right],
\end{equation}

\noindent where $\textup{S}_{n}$ is the power spectral density (PSD) characterising the detector noise {\qm and the bar denotes the complex conjugate}. Any template can now be normalised with respect to this inner product. For a template $x$, let the normalising factor {\qm $\langle x, x \rangle$} be labeled $c_{x}$. Taking the detector strain as being $s(t) = n(t) + h(t)$ where a signal $h$ is expected, with $n$ being the noise, the SNR $\rho_{s}(x)$ of the normalised template $x$ in the strain $s$ is defined as:

\begin{equation}
    \rho_{s}(x) := \langle s, x \rangle.
\end{equation}

In the presence of a signal, meaning $h$ is not identical to zero, the measured SNR {\qm (signified by a tilde)} $\tilde{\rho_{x}}(t)$ for a signal of amplitude $A$ is a random variable {\qm normally} distributed as $\mathcal{N}(c_{x}A, 1)$ \cite{Siemens, MFpart1, MFpart2}. Using this observation, the data can be match filtered against a set of waveform templates called a template bank. The template that best matches the data will produce the highest SNR.

Matched filtering has two major drawbacks. The first is the need for a template bank that sufficiently covers the parameter space which in general can be of high dimension, showcasing issues with scalability. The second is that, specifically for cosmic string searches, matched filtering is not robust to glitches, confusing the two classes due to their similar morphology. These points argue the case that it is worthwhile to explore alternatives to matched filtering for candidate detection in search pipelines. One natural choice is that of neural networks which in theory can address both drawbacks. From a theoretical point of view, it is interesting to note that work is being done towards the replication of matched filtering as neural networks \cite{Yan}. One could then make a case that neural networks can strictly improve on matched filtering.

\section{Methodology}\label{sec:methodology}

For the task of training convolutional neural networks on both the as-of-yet undetected cosmic string cusp signals and the per definition unpredictable glitches, a dataset incorporating advanced domain knowledge needs to be constructed. Once this data format is established, the network architecture is treated, along with the design decisions involved. Finally, the methodologies for a comparison to the state-of-the-art and making interpretations of the deep-learning model are described.

\subsection{Construction of the Dataset}\label{sec:dataset}

\begin{figure}[!]
    \includegraphics[width=1\columnwidth]{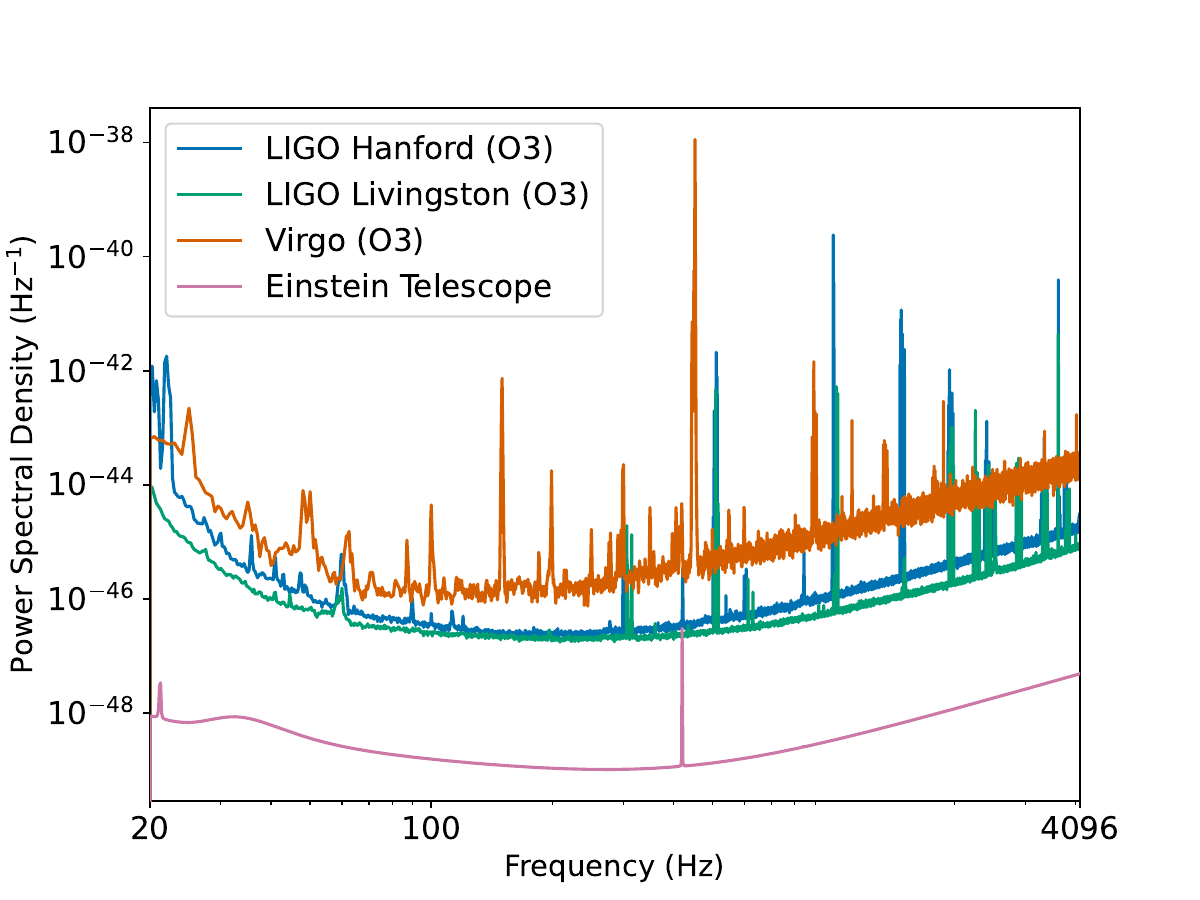}%
    \caption{The Einstein Telescope design sensitivity compared to the realised sensitivities of the current generation of detectors in the third observational run (O3).}
    \label{fig:psds}
\end{figure}

The Einstein Telescope will consist of three detectors in a triangular configuration \cite{etdesignreport}. As such, three detector strain data streams will simultaneously be collected. In this work, these streams will be labeled stream $0$ through $2$. Einstein Telescope data was simulated by first producing coloured Gaussian noise and then injecting cusp signals and blip glitches into the streams.

For each of the three streams of the Einstein Telescope, a Gaussian noise time series of length twelve seconds was generated, that was subsequently coloured by the PSD representing the Einstein Telescope design sensitivity \cite{ETASD} using \texttt{PyCBC} \cite{pyCBC}. The design sensitivity of the Einstein Telescope along with the sensitivities of current (second) generation detectors \cite{O3ASD} are shown in Fig. \ref{fig:psds}. The noise realisations were then injected with cusp signals to form the positive class, and artificial glitches to form the negative class.

\begin{figure}[!]
\includegraphics[width=1\columnwidth]{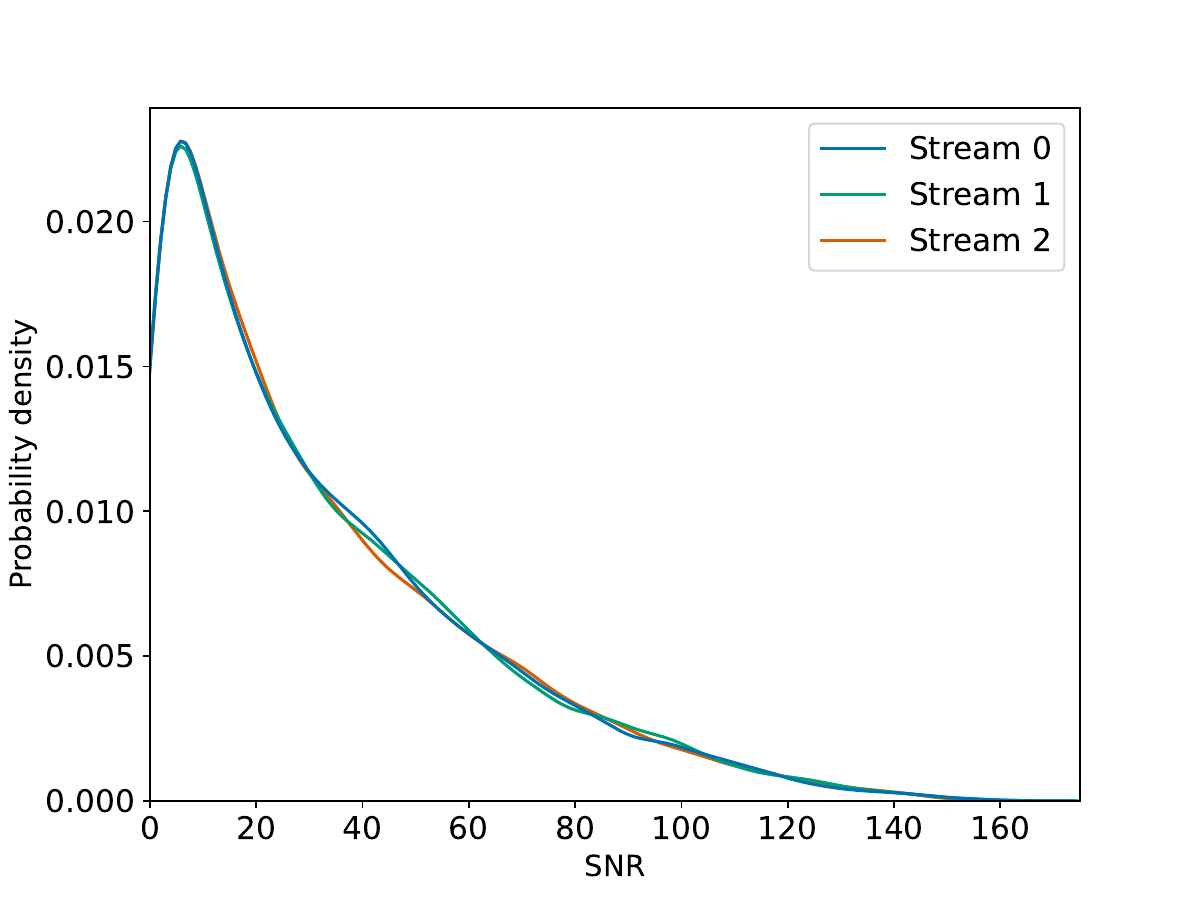}%
\caption{The SNR distribution of injected cusp waveforms {\qm modelled as a probability density.} The streams refer to the streams of the detectors making up the Einstein Telescope.}
\label{fig:inj_snr}
\end{figure}

The cusp waveforms in the time domain were generated through the use of the \texttt{LALSimulation} package \cite{lalsuite}. The function for the generation of the plus-polarised cusp strain components requires three inputs: an amplitude $A$ in $\textup{Hz}^{1/3}$ {\qm (normalising Eq. \ref{eq:cuspwaveform})}, a high-frequency cutoff $f_{\textup{high}}$ in $\textup{Hz}$ {\qm past which the waveform will drop exponentially}, and a sample period $\Delta_{t}$ in $\textup{Hz}$. Here, $A$ represents the amplitudal prefactor in Eq. \ref{eq:cuspwaveform}:

\begin{equation}
    A := 0.85 \frac{l^{2/3} G\mu}{(1+z)^{1/3} r(z)},
    \label{amplitudepre}
\end{equation}

\noindent so that the waveform in the time domain is given by the inverse Fourier transform of:

\begin{equation}
    h(f) = A f^{-4/3} (f_{\textup{high}} - f)^{+}
\end{equation}

\noindent where the plus signifies the taking of the positive part, and the output time series is of this transformed function at sample period $\Delta_{t}$. In order to randomly generate waveforms, the amplitudes {\qm $A$} and cutoff {\qm frequencies $f_{\textup{high}}$} were uniformly sampled from $[10^{-23}, 10^{-21}]$ and $\{20, 21, ..., 4000\}$ respectively. These generated cusp signals, assuming an isotropic distribution, were projected according to the Einstein Telescope antenna pattern and injected into Gaussian-coloured noise sampled at $8192$ Hz, before the strain was whitened and cropped to a length of eight seconds. The resulting SNR distributions of this positive class are shown in Fig. \ref{fig:inj_snr}.

The glitches were artificially generated using the \texttt{gengli} package \cite{gengliweb}, which has learned to model blip glitches in the time domain by harnessing generative adversarial networks \cite{Lopez1}. Currently, \texttt{gengli} approximates the real distribution of glitches in O2 data, specifically that of LIGO Hanford and Livingston. This data includes anomalies, and it is, therefore, possible anomalies showing a different morphology than blip glitches are generated. Using the \texttt{gengli} similarity metrics, an accepted region is defined that excludes roughly one in ten glitches that are deemed too dissimilar from blip glitches. These outliers are discarded. This procedure is described in \cite{Lopez2}.

The true morphology and intensity of Einstein Telescope glitches are currently unknown. It is however reasonable to assume that short-duration glitches similar to blips will be present in the recorded data, and as they are in fact a worst-case scenario in terms of similarity to the cusp signals, they form the best possible preparation. In order to further ensure the robustness of the models to be trained on this dataset, the generated glitches are scaled in amplitude to follow the SNR distribution of the injected cusps shown in Fig. \ref{fig:inj_snr}. {\qm This ensures that the models do not learn a difference in SNR distribution.} The injection procedure itself differs from that of cusp signals, since \texttt{gengli} generates whitened glitches. These glitches are summed as a time series to eight seconds of whitened noise at randomly drawn offsets. The offsets per stream are uncorrelated and the glitches are chosen randomly, meaning there is no detector coincidence for the glitches. Examples of both injected glitches and cusp signals are shown in Fig. \ref{fig:compsignal} and Fig. \ref{fig:compglitch}.

For both classes, no further preprocessing has taken place. In order to both preserve the original information and retain computational efficiency, time series are used instead of alternative representations like spectrograms.

The resulting dataset consists of 30,000 examples (or data points), split into training, validation and test sets of sizes 16,000, 4000 and 10,000 respectively. Each subset is balanced, meaning it is made up of equal parts positive examples (signals) and negative examples (glitches).

\subsection{The WaveNet Architecture}\label{subsec:wavenet}

The convolutional neural networks {\qm \cite{10.5555/1162264}} discussed in this section are implemented in \texttt{PyTorch} \cite{PyTorch} and were run on the LIGO Data Grid. The specific machine used has the following specifications: Intel E5-2670 CPU, NVIDIA Tesla V100 16GB GPU, and 128 GB of memory.

\begin{figure}[!]
    \includegraphics[width=1\columnwidth]{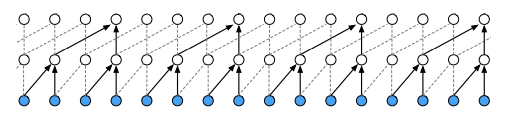}%
    \caption{Dilation between the layers {\qm of the neural network (shown horizontally)}, {\qm retrieved} from \cite{WaveNet}. As data is passed upwards through the layers, an increasing number of neurons is passed over, creating a larger diagonal reach for the neurons in the top layer.}
    \label{fig:dilations}
\end{figure}

\texttt{WaveNet} \cite{WaveNet} is an expressive convolutional neural network designed for the generation of high-fidelity speech audio. The architecture is capable of handling long-range temporal dependencies at high sampling rates, achieved by creating a large receptive field through the use of dilated convolutions, or dilations. Dilations allow the network this reception by skipping over a preset number of neurons in each layer, dilating the layers. By appending dilated layers, an exponential increase in the receptive field is gained at the cost of a linearly increasing number of layers, as is illustrated in Fig. \ref{fig:dilations}.

\begin{figure*}[htb]

\includegraphics[width=0.8\textwidth]{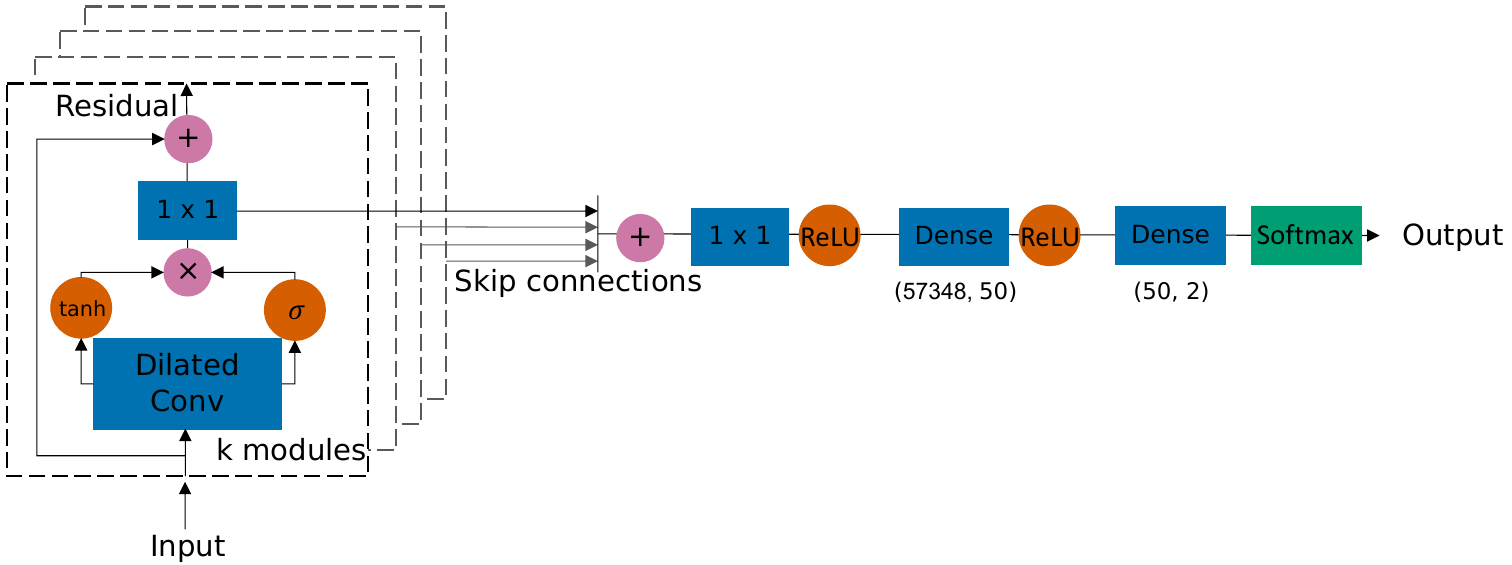}%
\caption{Overview of the modified \texttt{WaveNet} architecture for a single data stream, adapted from \cite{WaveNet}. {\qm The hidden layers are coloured blue, the internal activations are shown in orange, and the normalising softmax layer is shown in green.}} 
\label{fig:wavenet}
\end{figure*}

The major building blocks of \texttt{WaveNet} are residual block modules as presented in Fig. \ref{fig:wavenet}. The figure shows that input to the module is passed through a convolutional layer, after which it is simultaneously passed through both tanh and sigmoid gates. The activations {\qm \cite{10.5555/1162264}} are recombined in elementwise multiplication, where the sigmoid activations modulate the throughput, determining how much of the tanh output activation is passed \cite{PixelCNN}. The output is convolved with $1 \times 1$ filters to reduce the number of parameters before being fed into the residual connection \cite{Inception, ResNet}. Note that at this point a copy of the throughput is sent to a skip connection \cite{ResNet}.

Inspired by the methodology proposed in \cite{Wei:2020ztw}, where the full \texttt{WaveNet} architecture was modified for the discovery of binary black hole systems, modifications have been made for this project as well. The major changes are listed below.

\begin{itemize}
\item Instead of encoding the time series amplitudes in a range of $256$ possible values (see \cite{WaveNet} for details), no such limit is imposed in our implementation;
\item The causal structure intended for the dependencies in human speech was removed so as to provide the most possible information to the model;
\item The dilated convolutions have a kernel size of $3$ to capture fine details, and the dilation within the $k$-th block module (of $11$) is set to $2^{k}$;
\item The steps preceding the softmax activations were removed in favour of dense layers. In order to produce a probability, the activations need to be collapsed onto a scalar value in the unit interval. This too is shown in Fig. \ref{fig:wavenet}.
\end{itemize}

\noindent Together, these changes tailor the architecture to the needs of binary classification instead of the originally intended generation.

\subsection{Design and Parameter Choices}\label{sec:training}

The first major design choice is the use of an ensemble. Instead of training a single network on the three streams, one network was trained for each, and the three final networks were combined into an ensemble. This has several advantages. The first is the handling of different glitches being injected into the streams, therefore not allowing the ensemble to resort to using coincidence for its classification and forcing it to consider morphologies. Second, the independent networks can learn different characteristics during their training phase, averaging out to a more well-informed final decision by the ensemble. This average is taken literally, as the probability $\mathbb{P}(x)$ output by the ensemble for an example $x = (x_{0}, x_{1}, x_{2})$ {\qm of strains} is the average of the components networks $\mathbb{P}_{i}$ for $i \in \{0, 1, 2\}$:

\begin{equation}
    \mathbb{P}(x) = \frac{\mathbb{P}_{0}(x_{0}) + \mathbb{P}_{1}(x_{1}) + \mathbb{P}_{2}(x_{2})}{3}.
\end{equation}

\noindent When the time does arrive that coincidence is needed to confirm a candidate detection in joint analysis, these probabilities can be transferred to a central machine instead of the data containing the candidate. This greatly reduces latency, as a single probability is less costly to transmit than a time series.

The weights of each network were determined using stochastic gradient descent, specifically using the \texttt{AdamW} optimiser \cite{AdamW} with learning rate $10^{-4}$ and a weight decay of $10^{-3}$. These values were further varied, yielding no significant improvement at this small scale. The batch size was set to $13$. Due to the complexity of the model, increases in the batch size resulted in a direct gain in performance, and this trend is likely to continue. For this model, the batch size was limited by memory.

\begin{figure}[!]
    \includegraphics[width=\columnwidth]{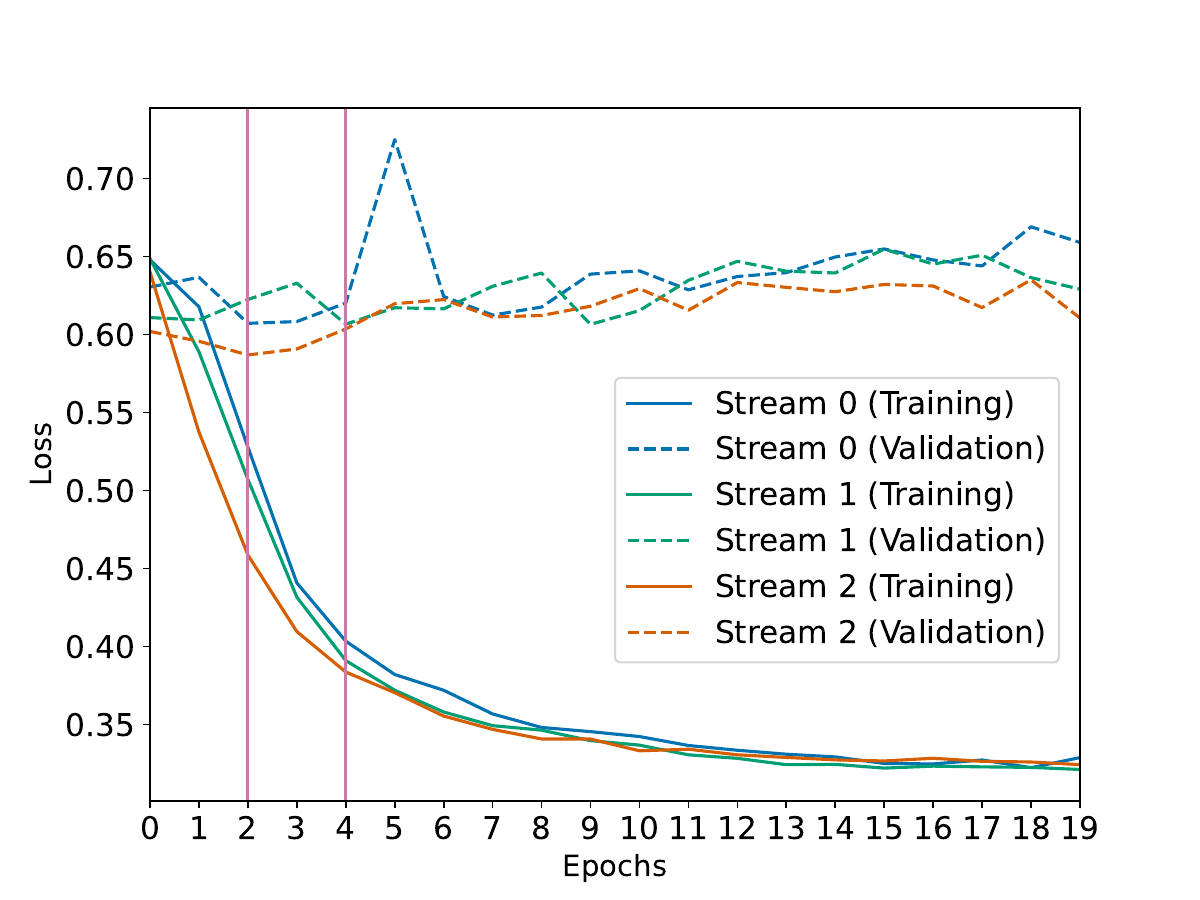}%
    \caption{Training and validation losses for the three networks. The stopping times are represented by vertical lines, marking the weights used for the networks.}
    \label{fig:losses}
\end{figure}

In the training phase, each separate network was trained independently for 20 epochs, resulting in the training and validation {\qm cross-entropy} losses shown in Fig. \ref{fig:losses}. This phase was repeated multiple times to ensure the optimiser did not get stuck in an avoidable local minimum. It can be read from the validation losses that because of the small batch size, overfitting started between the second and fourth epochs, marked by vertical lines. 

From here on, an ensemble is defined by the three ordered epochs at which the training of the networks was halted, denoting the ensemble so created as an {\qm $[i, j, k]$}-ensemble for {\qm $i, j, k$} between the values of $0$ and $19$. Choosing weights according to the times where overfitting started, the $[2, 4, 2]$-ensemble was established as the initial candidate. Classifiers defined by nearby stopping times in the {\qm $i, j, k$} lattice were checked by brute force iteration but gave no improvement over the $[2, 4, 2]$-ensemble. This ensemble was therefore chosen.

\begin{figure}[!]
    \includegraphics[width=1\columnwidth]{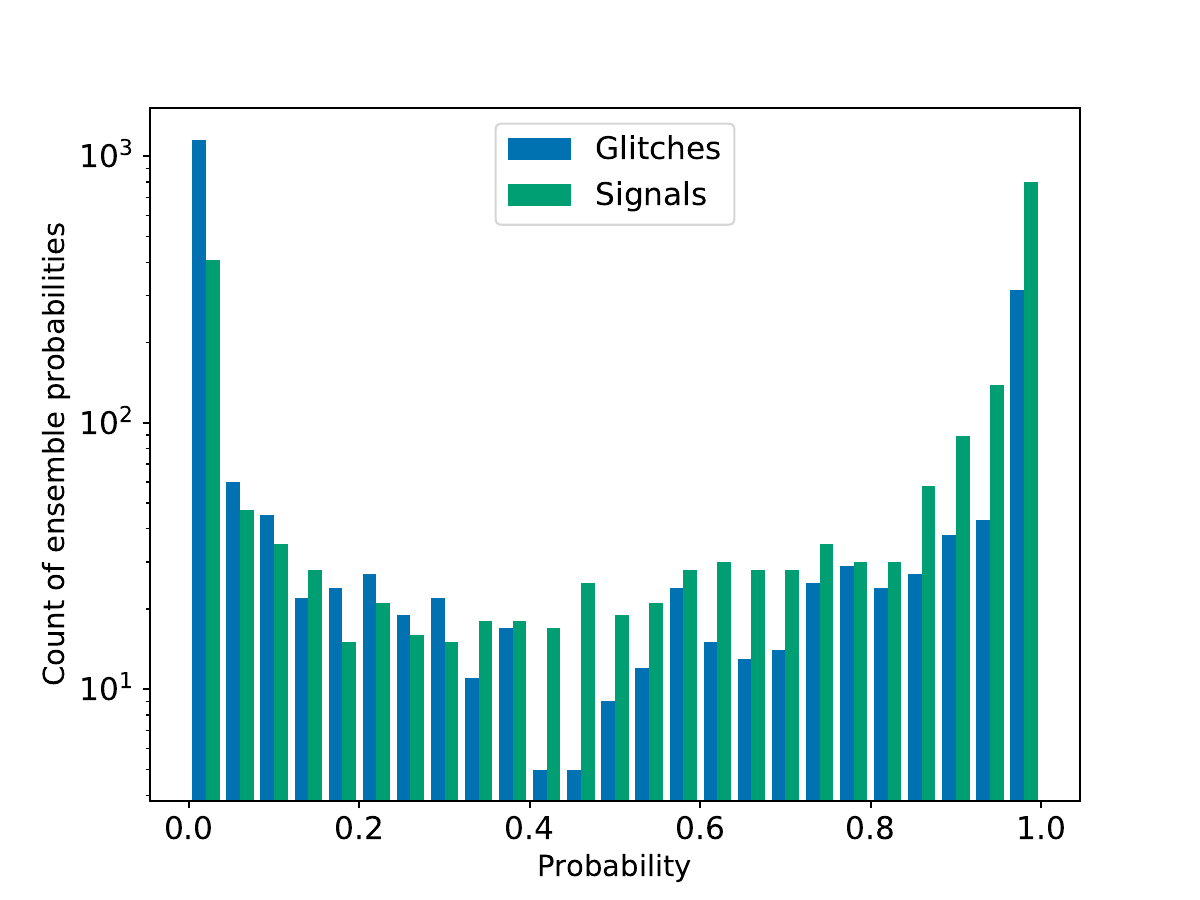}%
    \caption{Counts of the ensemble probabilities for both classes, with the y-axis in the log scale.}
    \label{fig:probabilities}
\end{figure}

Both the individual network thresholds, the ensemble threshold, and combinations of the two were fine-tuned on the validation set. The most important measures used in the fine-tuning were the accuracy, true positive rate and false positive rate. As can be seen from Fig. \ref{fig:probabilities}, the ensemble probabilities {\qm $\mathbb{P}$ on the validation set} are highly concentrated in the neighbourhoods of $0$ and $1$, giving no cause to deviate far from an ensemble threshold of $0.5$. The viable range for thresholds to test was set to the uniform set spanning from $0.4$ to $0.6$ with step size $0.01$, with none leading to a significant improvement over the default value of $0.5$. A similar line of reasoning has led to thresholds of $0.5$ for the component networks.

\subsection{Comparison to Matched Filtering}\label{subsec:method_matchfcomparison}

A direct comparison between the deep-learning model, a binary classifier, and matched filtering, which is not a binary classifier, is not straightforward. In order to benchmark the deep-learning model against matched filtering, a new balanced dataset of total size $400$ was constructed, again following the SNR distribution shown in Fig. \ref{fig:inj_snr} for both signals and glitches injected into Gaussian noise coloured by the Einstein Telescope design sensitivity. Recall that these were labeled the positive and negative examples respectively.

For a given positive example, each of the three streams was match filtered against the exact injected waveform. This waveform is per definition the optimum filter, and the trigger value is defined as the global maximum of the three SNR time series.

For the negative examples, the optimum filter will not exist as a cusp waveform template, as no true signal was injected. The choice of templates is therefore arbitrary. In order to simulate realistic circumstances, a template bank was created by randomly sampling $20$ of the $200$ signals generated during the creation of the comparison dataset. Including more templates would be detrimental to the performance of matched filtering, as these additional templates would only allow for the measured SNR to be increased, where it is known no cusp signal is present. Hence, the results from this comparison can be considered conservative. The performance of matched filtering could only be improved by constructing a template bank of cusp waveforms where no template can be matched to blip glitches, which defeats the purpose of the comparison. The remainder of the procedure is identical to that for the positive examples so that the method is internally consistent.

\subsection{Model Interpretability}\label{subsec:method_interpretability}

Neural networks are notoriously hard to interpret because of their large dimensionality and opaque optimisation procedures. Ideally, however, the discriminative properties the networks have learned would be extracted, in order to better understand the morphological differences between the injected signals and injected glitches. So as to learn what the neural networks have learned, a variety of methods is proposed to interpret the behaviour of our deep-learning model.

\subsubsection{Surgeries}\label{subsubsec:method.surgeries}

The first method employed to better the understanding of our deep-learning model is what will be referred to as glitch surgeries. Surgeries are limited to the class of glitches which are not subject to detector antenna patterns, meaning impact can more directly and accurately be measured for glitches. Moreover, they are more readily split into different regions on which surgeries can be performed. The predetermined parts of a selected glitch are excised before reclassifying the modified example and quantifying the change in the ensemble prediction with respect to the original input. In doing so, features salient to the deep-learning model can be identified.

\begin{figure}[!]
    \includegraphics[width=1\columnwidth]{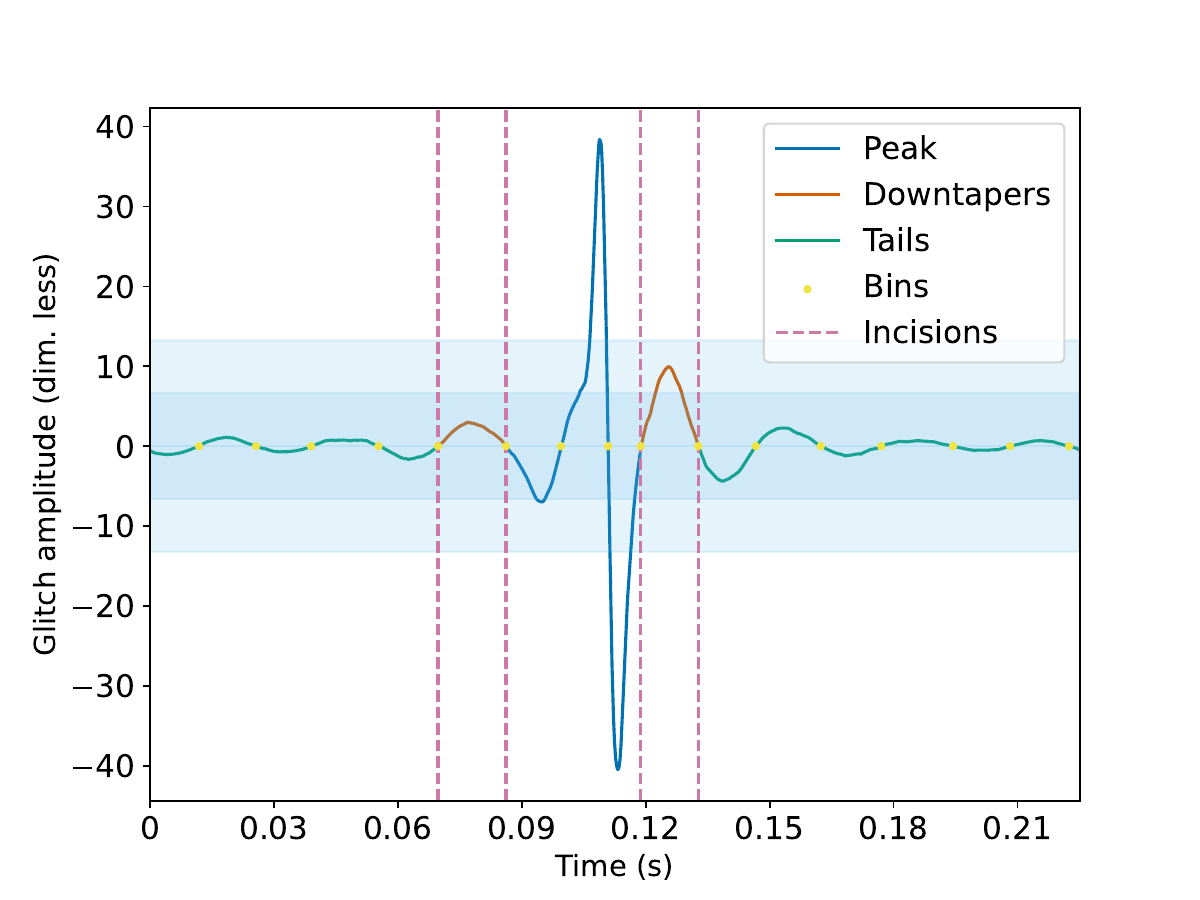}%
    \caption{The sections of a glitch within a visualisation of the surgery procedure. The shaded areas represent the standard deviation of the amplitude from zero.}
    \label{fig:glitch_surgery_zeroes}
\end{figure}

The observation underpinning the procedure is that a glitch {\qm $g(t)$} can generally be divided into five regions based on the maximum amplitude within these regions and that these regions together form the sections shown in Fig. \ref{fig:glitch_surgery_zeroes}. These regions can be automatically detected by partitioning a glitch into bins delimited by the zero crossings and comparing the absolute maximum of each bin with a function of the standard deviation $\sigma$ of the glitch amplitude. The edges of the bins are constrained to correspond with zero-crossings to ensure continuity, as an excision amounts to setting the value of the glitch waveform amplitude to zero within the bins that are excised. Whereas continuity is required so the neural networks do not pick up on the transitions, it is not necessary to extend the waveform to be smooth at the bin edges, as this transition is lost within the noise after injection.

First, the area of peak activity, which one should note may contain more than one peak, is identified as being the bin containing the absolute maximum amplitude $\lvert\textup{max}(g) \rvert$ of the glitch in the time domain. Moving outwards left and right over the bins, starting from the identified bin, the peak area is extended to include adjacent bins if the absolute maximum within these bins exceeds $2\sigma$. The downtaper of the glitch starts at the first bin where the absolute maximum within the bin is below $2\sigma$, and the tail of the glitch starts at the first bin where the absolute maximum is below $\sigma$. Note that these definitions may imply the absence of named sections in a glitch waveform, as for instance, a section corresponding to the downtaper might not exist. This can be the case if the maximum amplitude of the waveform is extremely high compared to the average amplitude. Such glitches can safely be included in the surgery procedure. The excision of a non-existent section amounts to nothing changing at all, and the results from the reclassification will reiterate that the non-existent section did not contribute to the classification.

For a representative sample taken from the dataset, the procedure is then as follows.

\begin{enumerate}
    \item Choose a glitch example $g$ from the set, and retrieve the ensemble probability $\mathbb{P}(g)$;
    \item For any of the three sections, set $g$ identical to zero within the corresponding bins to obtain $g_{\textup{section}}$ (performing the surgery);
    \item Reinject and classify $g_{\textup{section}}$ before retrieving the ensemble probability $\mathbb{P}(g_{\textup{section}})$.
\end{enumerate}

\noindent The statistic of interest is then:

\begin{equation}
    \Delta_{\textup{section}}(g) := \mathbb{P}(g) - \mathbb{P}(g_{\textup{section}}).
\end{equation}

\begin{figure}[!]
    \includegraphics[width=1\columnwidth]{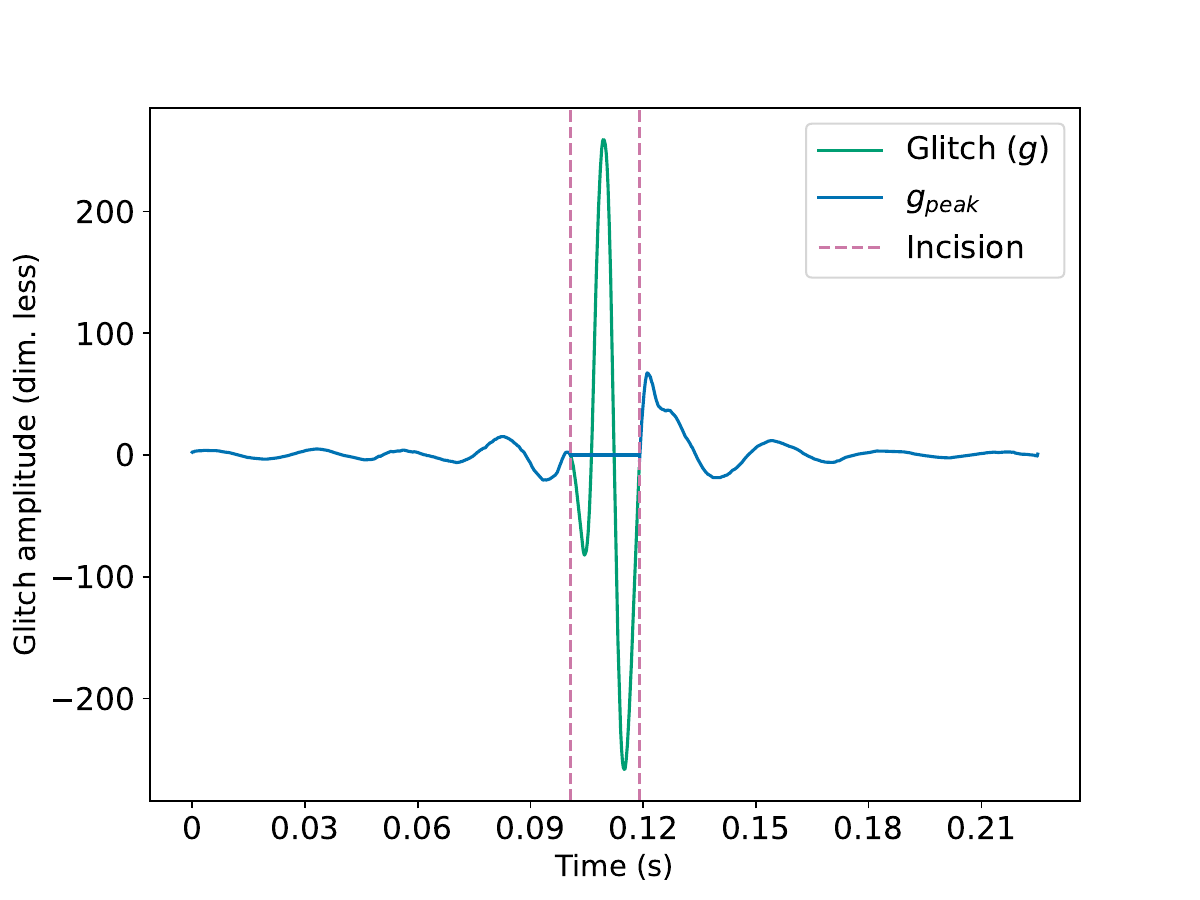}%
    \caption{A glitch before and after surgery. During the surgery the peak was excised from $g$, thus obtaining $g_{\textup{peak}}$. The corresponding statistic is $\Delta_{\textup{peak}}(g) \approx -0.47$.}
    \label{fig:glitch_surgery_excised}
\end{figure}

\noindent Note that this statistic takes values in $[-1, 1]$. The natural interpretation is that a value close to $-1$ means that the classification has significantly changed, with $g$ being classified as a glitch previously and as a signal following the surgery. A value close to $1$ would imply the reverse. A stream from an example with a value of $\Delta_{\textup{peak}} \approx -0.47$ is shown in Fig. \ref{fig:glitch_surgery_excised}. This behaviour can be further explored by considering the changes for the individual component networks within the ensemble.

\subsubsection{Activations}\label{subsubsec:method.activations}

Another way of investigating the behaviour of the ensemble is the extraction and visualisation of the activations in the hidden layers as a testing example is passed through. This information can then be used to tie certain convolutional filters to specific confusion matrix classes {\qm (Fig. \ref{fig:confusionmatrix})} in the dataset. Note that these are filters according to the terminology of neural networks, not those of matched filtering. Inspiration was drawn from saliency maps \cite{saliency} from computer vision, meant to highlight the most salient and therefore recognisable regions of images. Although projects like \texttt{Captum} \cite{captum} offer similar ways of interpreting convolutional neural networks, they differ from the method described here, designed specifically for the analysis of time series data.

A straightforward way of obtaining the activations (that also works for general networks) is to deconstruct a given network into an ordered set of individual layers, applying these layers one by one, and saving the outputs before feeding the output forward. 
Once these values are recovered, the challenge of interpreting the activations is reduced to trying to connect the activation of specific filters to fundamental characteristics of the example that was passed through. This is akin to detailing a collection of neurons that fire when a specific example is seen. As this is an extremely difficult task with high dimensionality, only isolated observations can be made.

In order to understand how the activations can be best visualised, it is useful to review the process of a filter being applied. As a filter is convolved with the one-dimensional time series, a new time series containing a large number of activation values is obtained and fed out. Due to the number of values, a direct plot of the activations would be unreadable. Instead, the values are binned and smoothed with a kernel density estimate. The resulting curve is an indicative visualisation of the activation for the specific filter used. The reader is invited to look ahead at Fig. \ref{fig:activations}, which shows these curves for the examples that will be interpreted in the next section.

\subsubsection{Principal Component Analysis}\label{subsubsec:method.pca}

Whereas the extraction of the activations serves mostly to delve into the hidden representations of the data as it is passed through the modules, principal component analysis (or PCA) \cite{PCA} can be used to analyse the representation in the linear layers. PCA is a dimensionality reduction method that linearly maps vector data into a lower-dimensional space with an ordered basis consisting of what are called the principal components. These components are determined as being the basis vectors carrying the most amount of information measured by variance, and their ordering is based on these same amounts. This means that for instance, the first principal component contributes the most to the overall variance of the dataset. PCA is applied to the second to last dense layer shown in Fig. \ref{fig:wavenet}, where an input of size $57,348$ is collapsed to an output of size $50$ before the latter values are further reduced to a single probability. Based on these numbers one can argue that in this layer the most amount of information is condensed, making it a valuable object of study. In Sec. \ref{subsubsec:results.pca} the first two principal components obtained from the dense layer are studied.

\section{Results}\label{sec:results}

In this section, the numerical results of the chosen deep-learning model are reported and discussed before treating the information extracted from the model by applying the interpretability methods presented in Sec. \ref{sec:methodology}.

\subsection{Numerical Results}\label{sec:numresults}

\begin{figure}[!]
    \includegraphics[width=1\columnwidth]{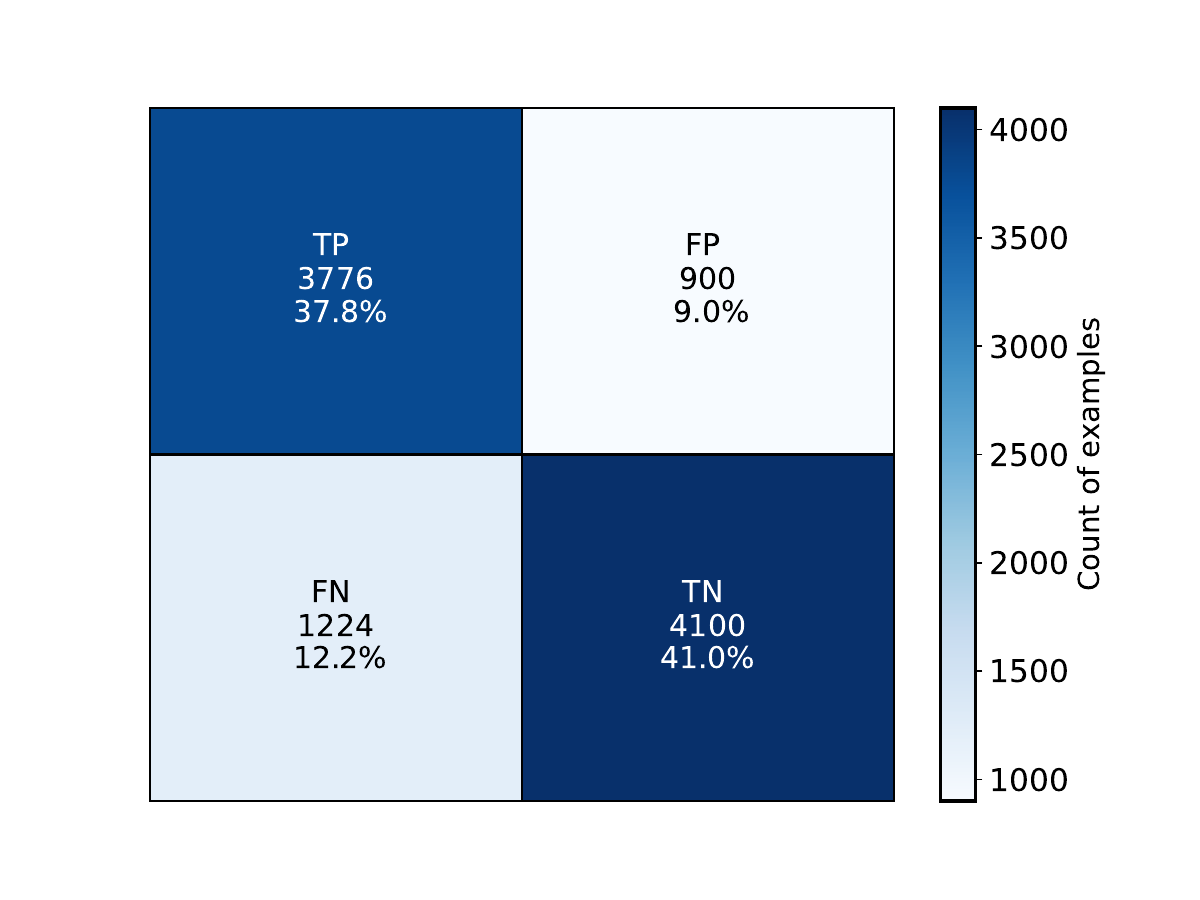}%
    \caption{{\qm The confusion matrix showing the true positives (TP), false positives (FP), false negatives (FN) and true negatives (TN) for the $[2, 4, 2]$-ensemble on the test set, visualised as a heatmap.}}
    \label{fig:confusionmatrix}
\end{figure}

\begin{table}[!]
	\begin{tabular}{|c c c|}
		\hline
		\textbf{Metric} & \textbf{Formula} & \textbf{Value} \\
		\hline
		Accuracy & \quad\quad (TP + TN) / (P + N) \quad\quad & 0.7876 \\
		\hline
		True Positive Rate & TP / (TP + FN) & 0.7552 \\
		\hline
		False Positive Rate & FP / (FP + TN) & 0.1800 \\
		\hline
	\end{tabular}
	\caption{A selection of performance metrics for the $[2, 4, 2]$-ensemble on the test set, along with their formulae. These values were computed from the confusion matrix in Fig. \ref{fig:confusionmatrix} {\qm using the true positives (TP), false positives (FP), false negatives (FN), true negatives (TN), with P := TP + FP (positives) and N := TN + FN (negatives)}.}
	\label{tab:metrics}
\end{table}

On the test set, the $[2, 4, 2]$-ensemble yields the confusion matrix visualised in Fig. \ref{fig:confusionmatrix}, from which the metrics presented in Table \ref{tab:metrics} were computed. In terms of cosmic string searches, the accuracy refers to the model's capability of recovering the injected signals and glitches. Likewise, the true positive rate (TPR) quantifies how well the model can recognise a signal, given that a signal was injected. Finally, the false positive rate (FPR) measures to what degree the model mistakes glitches for signals, given that no signal was injected. This means that a value of $0$ is ideal for the FPR, and $1$ is ideal for the accuracy and TPR.

\begin{figure}[!]
    \includegraphics[width=1\columnwidth]{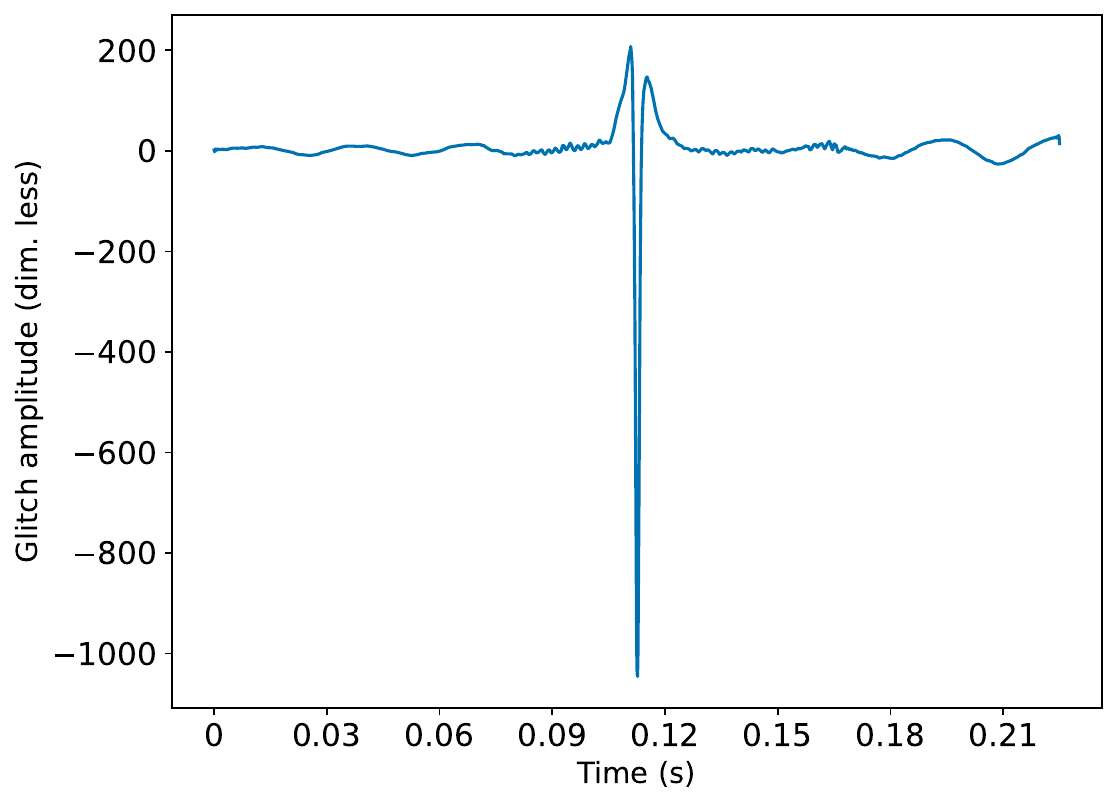}%
    \caption{An example of a glitch that was falsely classified as a positive with high probability. It is therefore an extreme example of a false positive.}
    \label{fig:fpexample}
\end{figure}

Manually investigating the most extreme false positive examples, spurred by the relatively high FPR, most skew high on this metric due to the three streams having been injected with glitches sharing a similar morphology between them, shown in Fig. \ref{fig:fpexample}. It appears the extremely high maximum amplitude (as compared to the average amplitude) dominates the morphology, leaving the deep-learning model little other distinguishing features to base its classification on. For these specific examples, the model erroneously resorted to a positive classification.

\begin{figure}[!]
    \includegraphics[width=1\columnwidth]{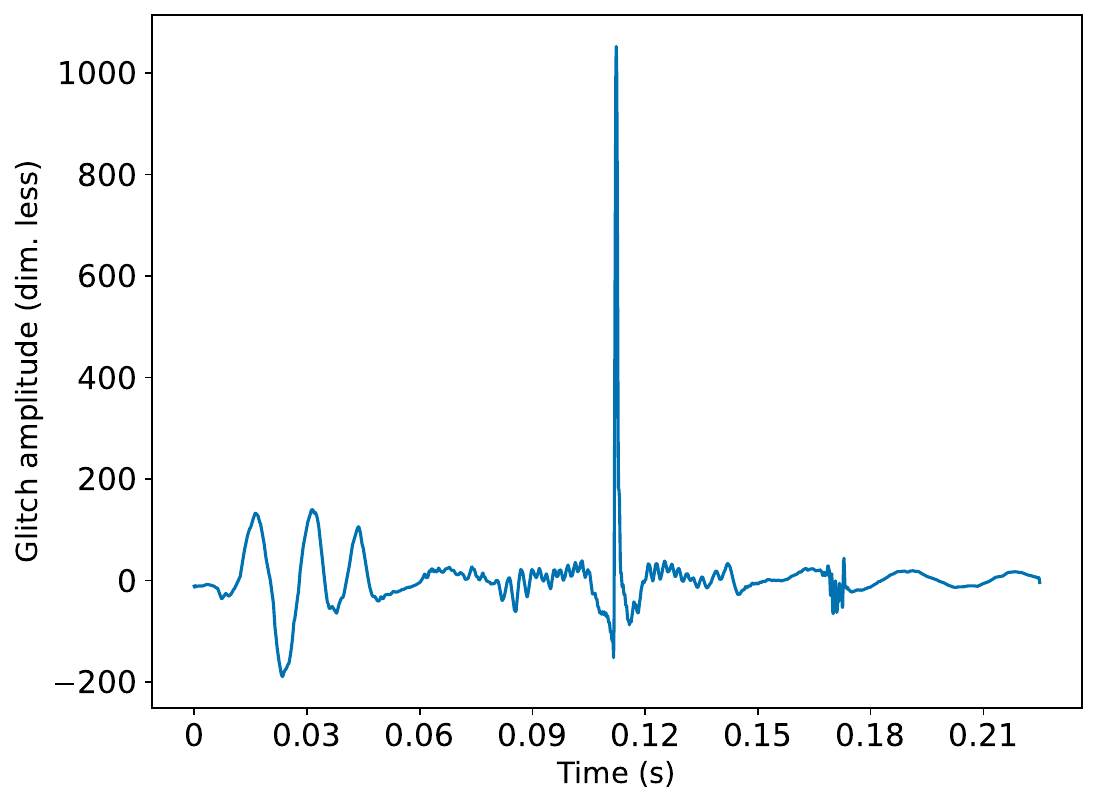}%
    \caption{A different extreme example of a false positive. One possible explanation for this misclassification is that the component network corresponding to this stream is given more opportunity to identify a signature that is believed to be that of a signal.}
    \label{fig:fpexample2}
\end{figure}

There are a few noteworthy exceptions, each appearing only in one of the three streams that make up an example, showing oscillations in amplitude over a larger period of time. Such a glitch is shown in Fig. \ref{fig:fpexample2}. One possible explanation for these instances is that the component network is given more opportunity to detect the presence of a signal signature and that one such signature is sufficient for the example to be classified as a signal. This underlines the importance of the signal morphology to the deep-learning model.

For the archetypal examples shown in Fig. \ref{fig:compsignal} and Fig. \ref{fig:compglitch}, the component networks for the streams these examples were taken from assigned the glitch a probability of $0.0008$ of being a signal, and the signal a probability of $0.7005$. This means the networks assign these examples to the right classes with considerable confidence.

Lastly, on the machine used (described in Sec. \ref{subsec:wavenet}), the classification speed of one example (consisting of three data streams of $8$ seconds) was computed to be $10$ milliseconds on average.

\subsection{Comparison to Matched Filtering}\label{subsec:results_matchfcomparison}

\begin{figure}[!]
    \includegraphics[width=1\columnwidth]{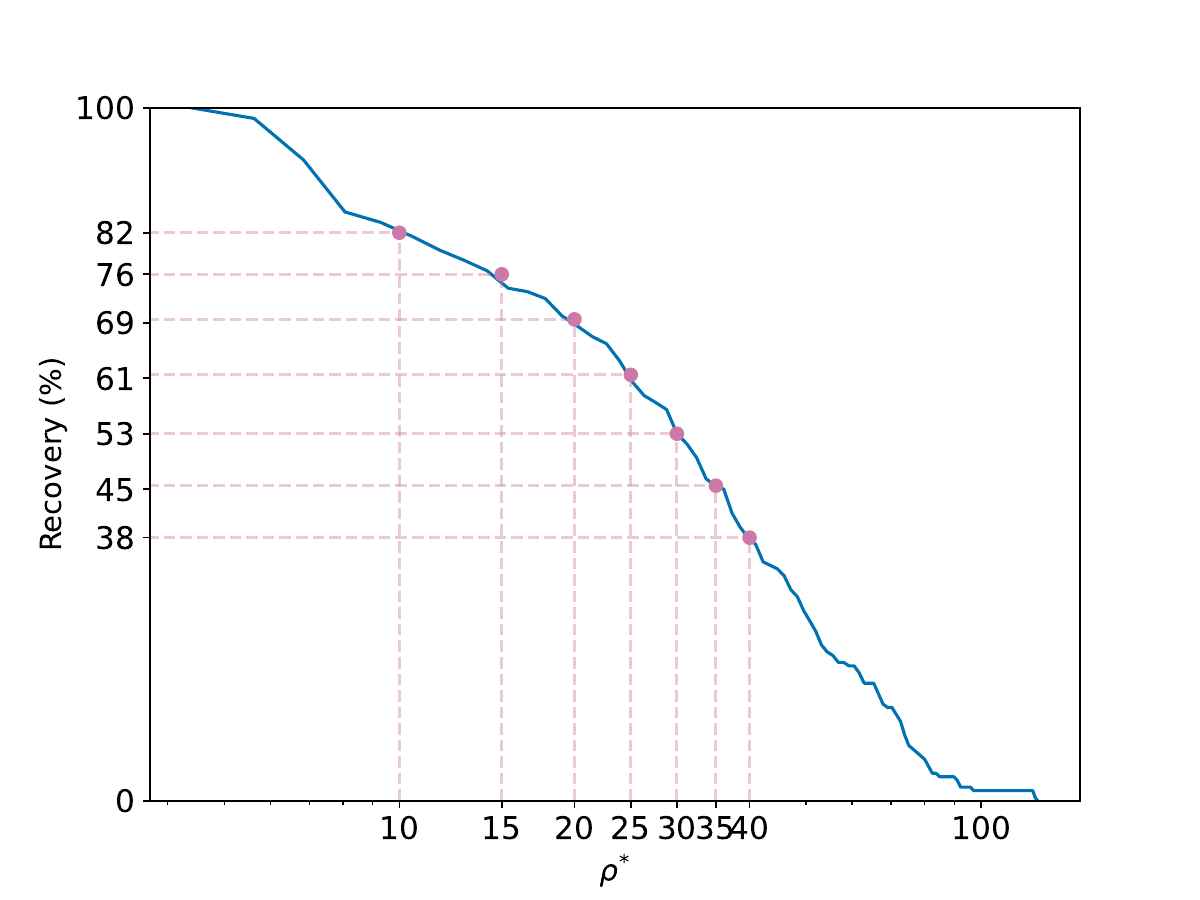}%
    \caption{The recovery percentage of matched filtering for different values of the SNR threshold $\rho^{*}$ in log scale. This latter value represents the cutoff from which point onwards an SNR is considered high enough for the corresponding example to be labeled as containing a signal.}
    \label{fig:recovery}
\end{figure}

\begin{figure}[!]
    \includegraphics[width=1\columnwidth]{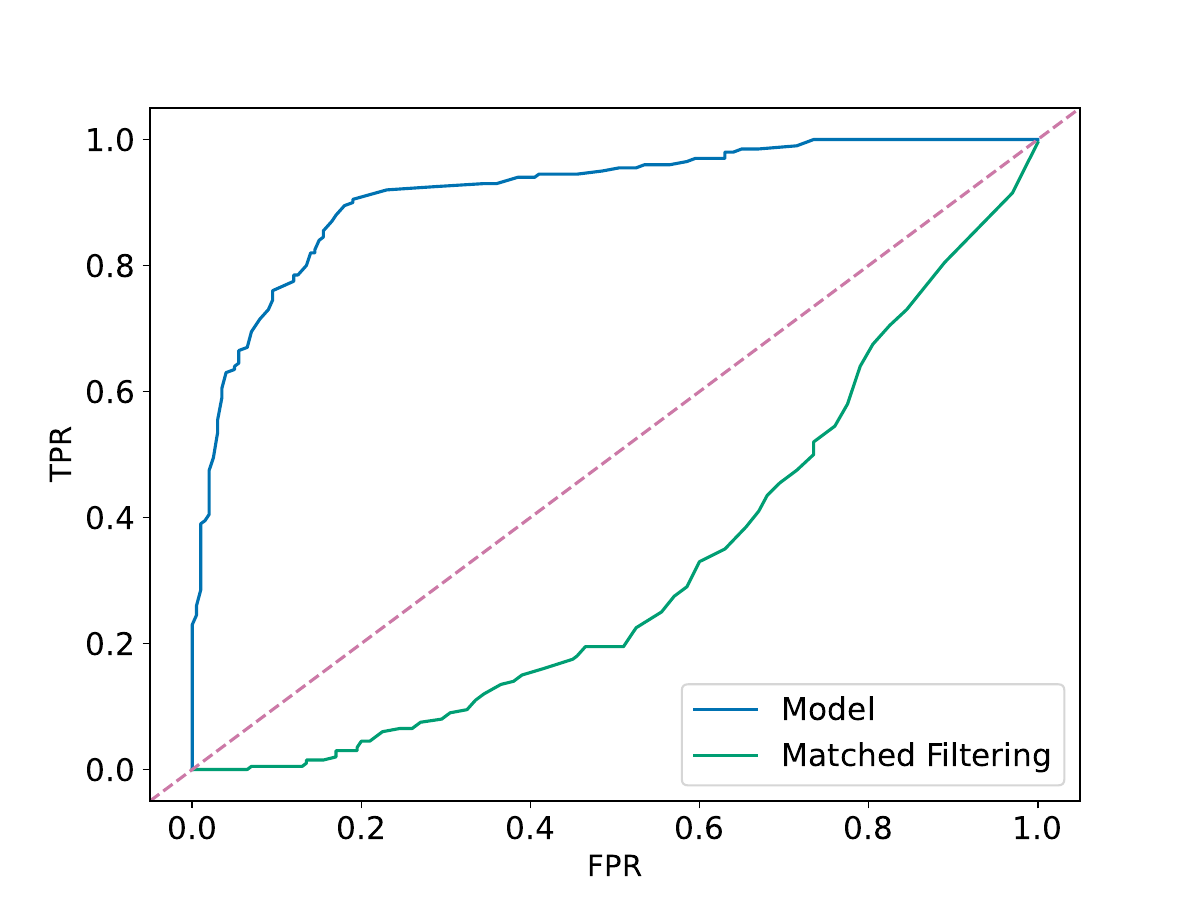}%
    \caption{The Receiver Operating Characteristic curves for both the deep-learning model and matched filtering on a dataset consisting of injected signals and glitches. The performance of a random binary classifier is represented by the diagonal.}
    \label{fig:roc}
\end{figure}

Conventionally, the performance of matched filtering is measured with positive examples being signals added to noise, and negative examples being drawn from a coloured Gaussian noise background without a signal present. Recall however that the current consideration is the distinguishing power of the deep-learning model and matched filtering for a dataset where the positive examples are injected signals, and the negative examples are injected glitches.

The true positives recovered by matched filtering at different SNR thresholds $\rho^{*}$ is shown in Fig. \ref{fig:recovery}. This choice for the representation of the results was made in order to remain agnostic towards the chosen threshold, which may differ per analysis. A direct comparison between the deep-learning model and matched filtering is shown through the Receiver Operating Characteristic (ROC) curves in Fig. \ref{fig:roc}. The diagonal represents a random binary classifier, meaning that on this dataset, matched filtering is weighed down by false positives to the point where its performance is worse than random classification. In contrast, the deep-learning model is very effective on the same dataset. The conclusion is that the model is better at rejecting glitches than a simple matched-filter search with cosmic string cusp templates by a large margin. A more complete comparison, including for instance the additional mechanisms that would be present in a full gravitational-wave search pipeline and would work to ameliorate false alarms, is deferred to future work.

\subsection{Interpretability}\label{sec:interpretation}

In this section the results of the various interpretability methods described in Sec. \ref{subsec:method_interpretability} for the interpretation of the deep-learning model are presented.

\subsubsection{Surgeries}\label{subsubsec:results.surgeries}

\begin{table}[!]
	\begin{tabular}{|c c c|}
		\hline
		\textbf{Section} & \textbf{Average}($\Delta_{\textup{section}}$) & \textbf{Maximum}($\Delta_{\textup{section}}$) \\
		\hline
		Peak & 0.161 & 0.832 \\
		\hline
		Downtapers & 0.001 & 0.055 \\
		\hline
            Tails & 0.004 & 0.205 \\
		\hline
	\end{tabular}
	\caption{The average and maximum values of $\Delta_{\textup{section}}$ statistics, separated per section. These values offer a summary of the output from the surgery process, where $\Delta_{\textup{section}}$ measures the impact of a glitch section removal on the model classification.}
	\label{tab:surgerydeltas}
\end{table}

By performing the surgeries described in Sect. \ref{subsubsec:method.surgeries}, the statistics given in Table \ref{tab:surgerydeltas} were obtained. These statistics show that with a division of a glitch into three sections, the peaks will by far be the most informative, with the downtapers and tails contributing relatively little. The low average and maximum values for the latter section statistics bound the values on the whole set of glitches, indicating that for no negative example the removal of either the downtapers or tails has made a significant impact on the reclassification.

Investigating the outliers near the maximum of $0.205$ for $\Delta_{\textup{tails}}$, it was found that these values stem from glitches with high fluctuations in tails that were removed. For the values on the low end of $\Delta_{\textup{peak}}$, a similar observation is made. The removal of the peaks for these glitches left behind fluctuations in amplitude in the downtapers or tails, on which the deep-learning model will presumably base its classification instead.
Most of the examples with a high value of $\Delta_{\textup{peak}}$ have very large amplitudes in the peak section, the removal of which confuses the model. An interesting note is that for these examples the network trained on stream $0$ seems to be less impacted by the excision of peaks than the other two networks in the ensemble, which is possible evidence that the three networks have learned to identify different glitch signatures. Further manual inspection of the small number of examples with $\Delta_{\textup{peak}}$ near $-1$, meaning the classification has changed from a glitch to a signal following the surgeries, shows that all have remaining fluctuations in their waveforms. One theory is that the model considers these remnants as the new peak sections, viewing at least one as evidence of a present signal. This observation might suggest that without detecting a clear glitch signature, the model defaults to a signal classification. This would complement the discussion on the false positives in Sec. \ref{sec:numresults}.

Relating to the preceding discussion, if the model indeed resorts to analysing amplitude spikes within the sections that remain after a surgery has been performed, this suggests the model considered these sections as secondary to the peak region before. In turn, this suggests that the model does not simply detect rapid changes in amplitude, but has learned to differentiate morphologies.

\subsubsection{Activations}\label{subsubsec:results.activations}

\begin{figure*}[!]

\subfloat[{True Negative ($\mathbb{P} = 0.006$)}]{%
\includegraphics[width=0.50\columnwidth]{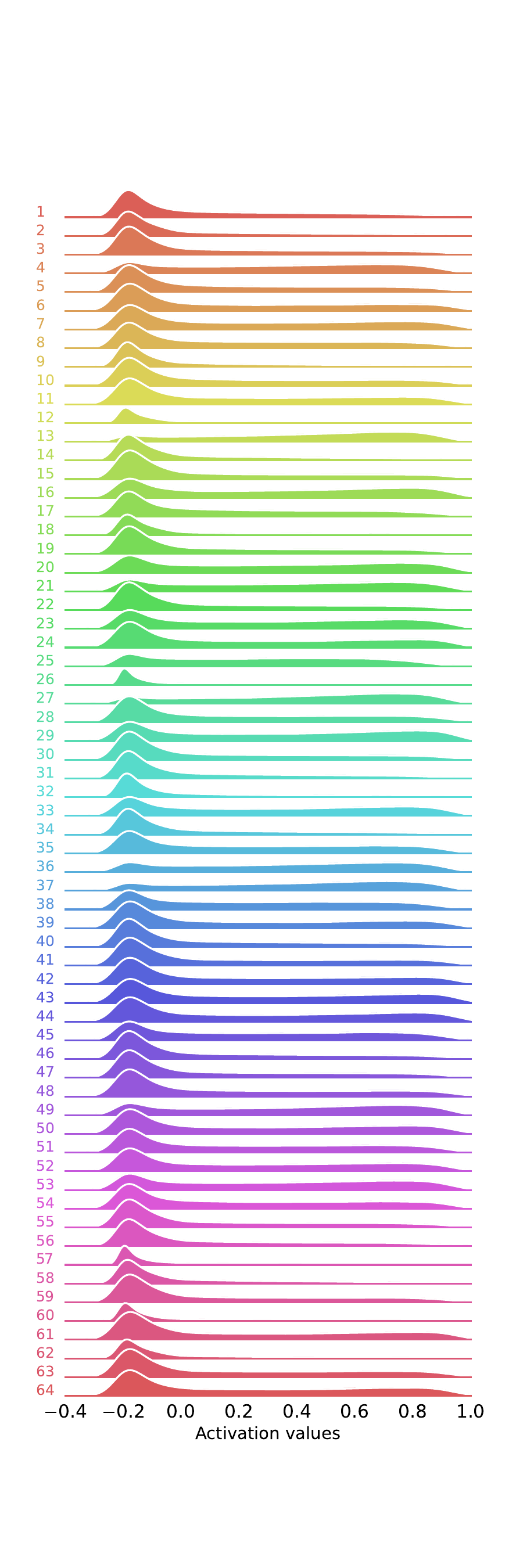}%
\label{fig:activations_tn}
}\hfill
\subfloat[{False Negative ($\mathbb{P} = 0.001$)}]{%
\includegraphics[width=0.50\columnwidth]{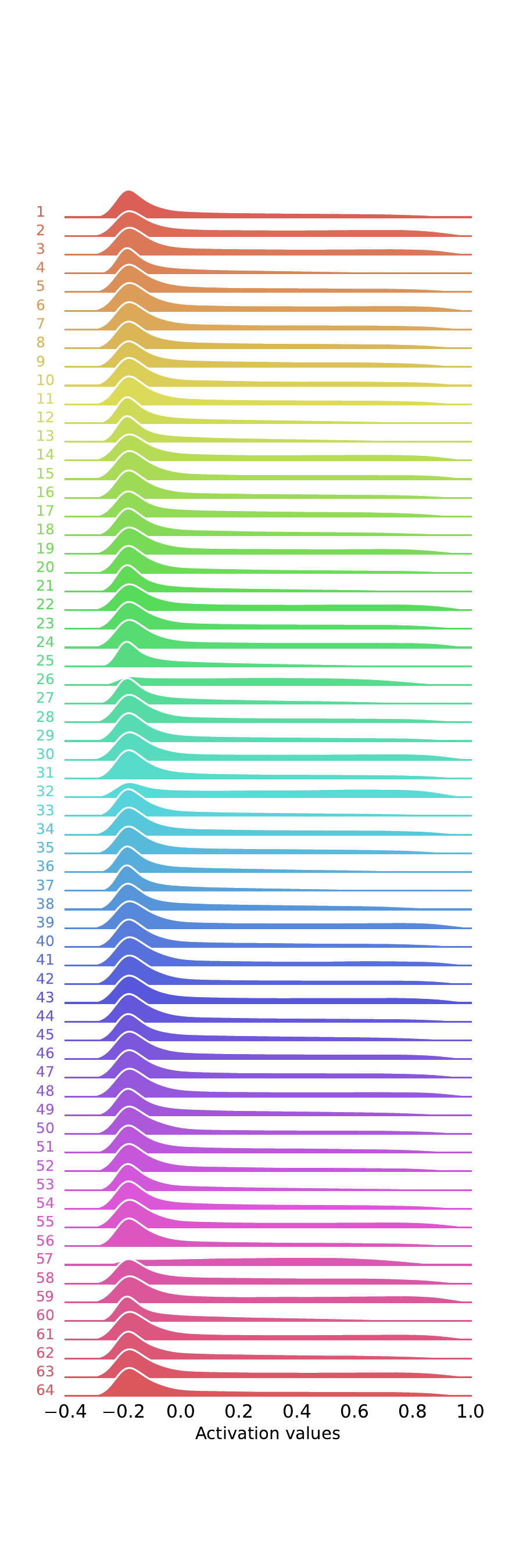}%
\label{fig:activations_fn}
}
\subfloat[{False Positive ($\mathbb{P} = 0.982$)}]{%
\includegraphics[width=0.50\columnwidth]{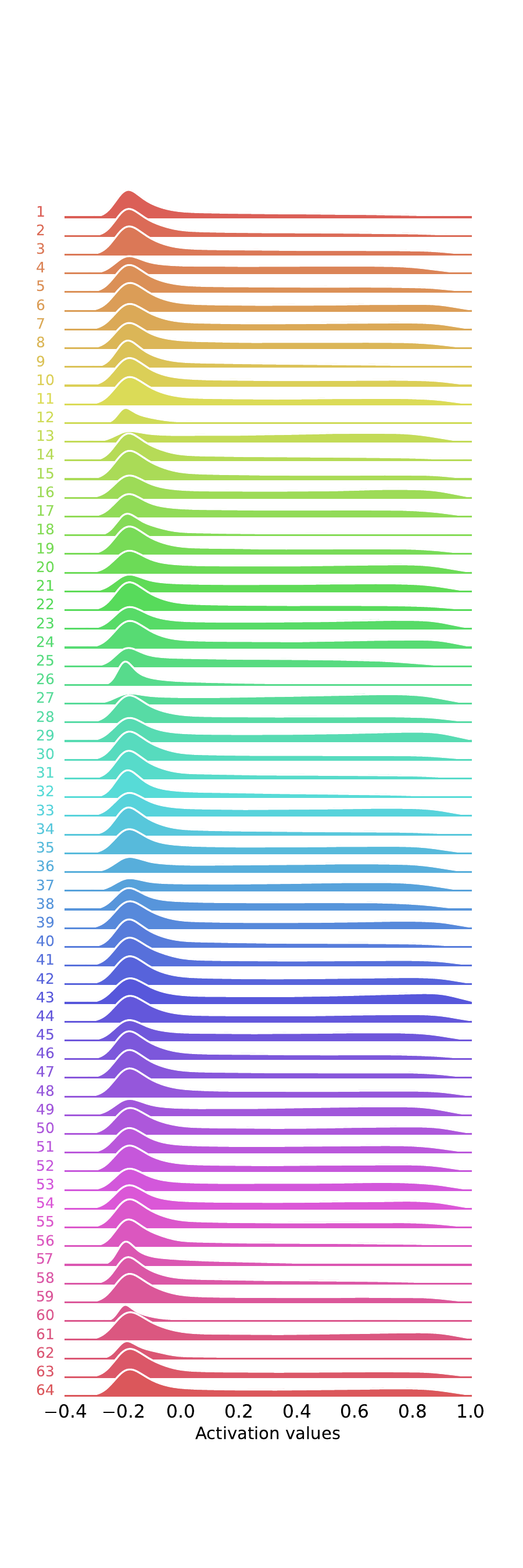}%
\label{fig:activations_fp}
}\hfill
\subfloat[{True Positive ($\mathbb{P} = 0.967$)}]{%
\includegraphics[width=0.50\columnwidth]{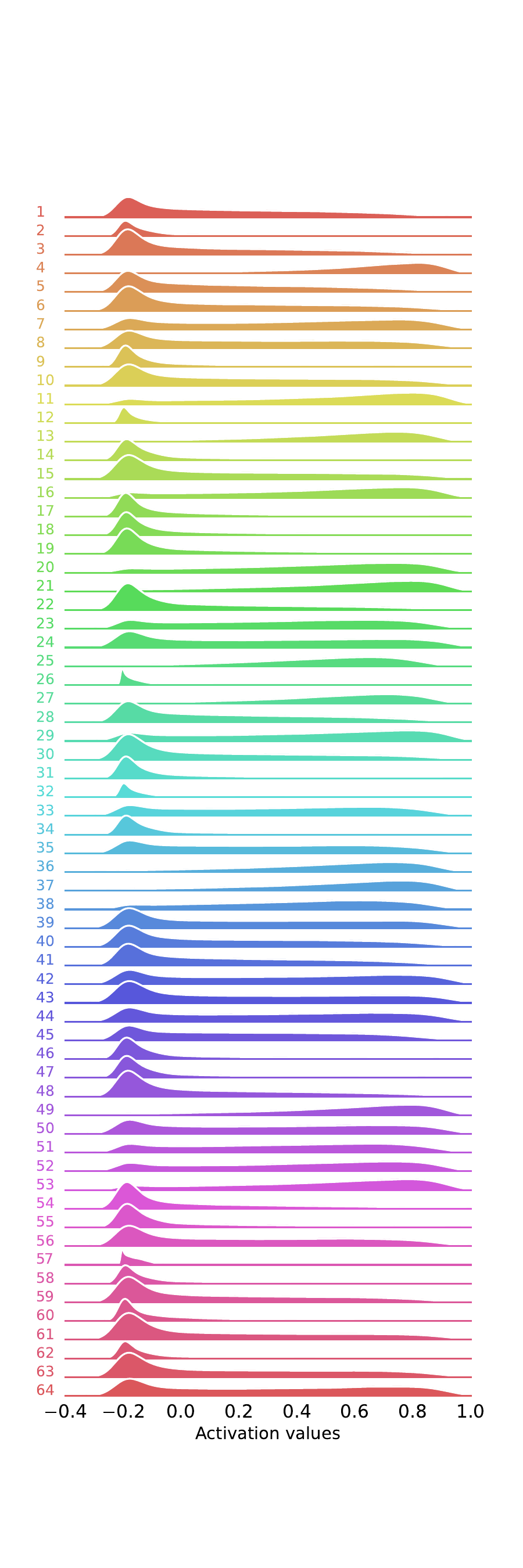}% 
\label{fig:activations_tp}
}
\caption{Residual activations {\qm output by the} $11$-th module in Fig. \ref{fig:wavenet}. Every horizontal axis represents a filter with the y-axis limited to $[0, 0.1]$, showing the distribution of activation values as a density curve. The mode of such a curve being close to $1$ indicates high activation values for the filter.}
\label{fig:activations}
\end{figure*}

Based on the confidence of the deep-learning model, one example was chosen for each of the classes in the confusion matrix, and their activation values were extracted. This means, for instance, that in the case of the true negative, an example with an output probability very close to $0$ was selected. The activations of these four examples for a single module are visualised in Fig. \ref{fig:activations} and each example will be discussed individually.

For the true negative example in Fig. \ref{fig:activations_tn}, a number of filters show activation, meaning the mean of the density curve is closer to $1$ than it is to $0$. However, this is with a high spread in the curve, indicating uncertainty in the activation values for this filter. Some filters, such as $27$ in green or $37$ in blue, show higher certainty. This is however not enough to mislead the model into making a false positive classification.

The filters in the false negative in Fig. \ref{fig:activations_fn} see barely any activation taking place at all. For this example, there was nothing giving the model the impression there could be a signal present. The only filters showing a semblance of activation do so with little certainty, with output probabilities not high enough to cross the classification thresholds {\qm at which point the model would classify the example as positive.}

The false positive example shown in Fig. \ref{fig:activations_fp} seems to invoke response from the filters, with activity within multiple filters. However, as was the case for the false negative, there is not much certainty. In this case, however, the probabilities did cross the classification thresholds.

Finally, the centroids of the true positive example in Fig. \ref{fig:activations_tp} skew strongly towards the right, meaning high values of the activations are achieved. A wide array of filters show strong activation, with certainty higher than the previous examples. The individual filters, and the network as a whole, are certain this example contains a signal.

The above observations were made on single examples, and are therefore not guaranteed to generalise. They do however show a clear difference in response to examples from the four classes and are therefore a proof of concept for further investigation. Individual filters can for instance be mapped back to certain sections of the input streams and examined further. This is outside the scope of this work.

As a final remark for this subsection, there are some filters that show little to no activation for any of the four examples, with $9$ in orange and $60, 62$ in red being such filters. While this is possibly due to the choice of examples or a lack of need for these filters, it is also possible this is a result of the low number of training epochs, meaning the weights for these filters have not been properly adjusted. If this is the case, one might conclude there is room to improve the model further. One of the ways this could be done is by reducing memory usage during the training phase, leading to better training that may in turn recruit the now dormant filters.

\subsubsection{Principal Component Analysis}\label{subsubsec:results.pca}

\begin{figure}[!]
    \includegraphics[width=1.01\columnwidth]{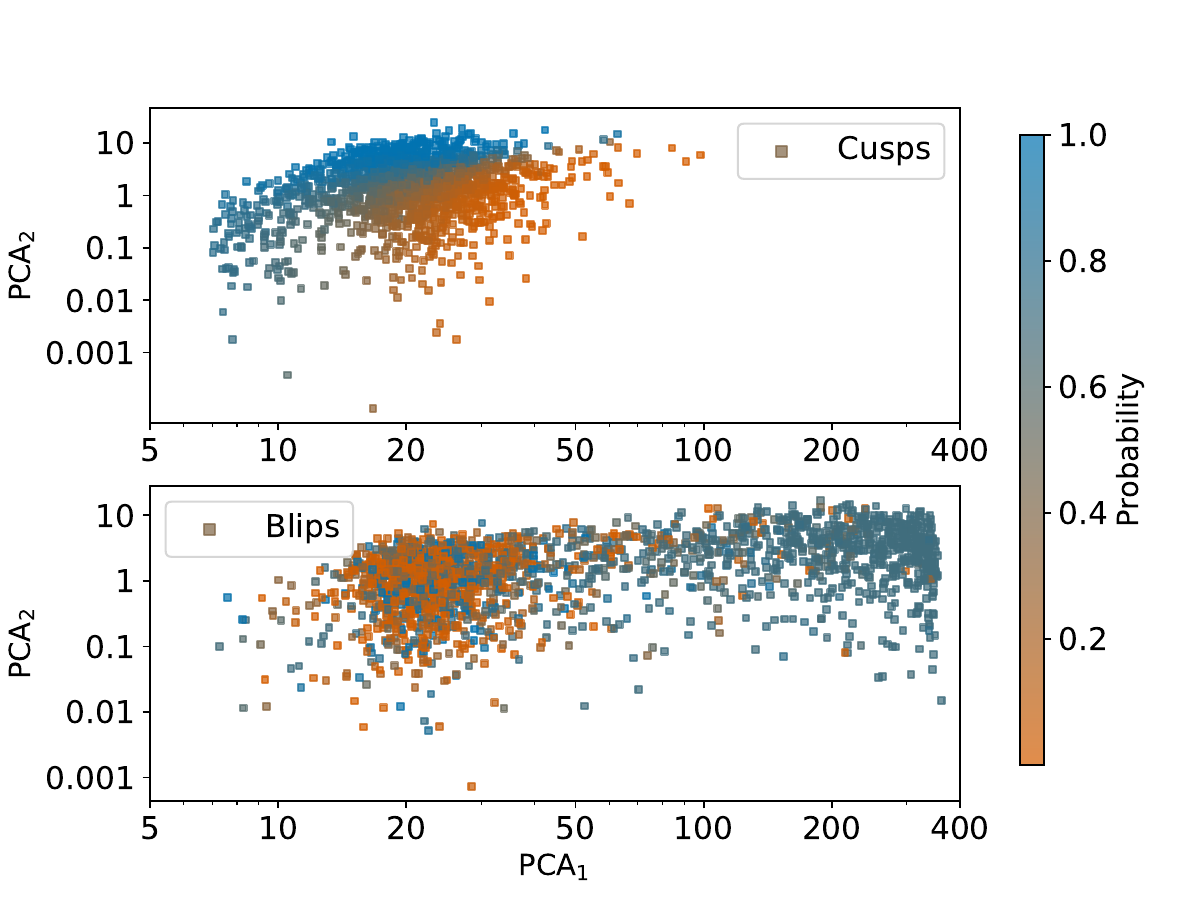}%
    \caption{A representation of a selection of activations from the dense layer in the space spanned by the first two principal components {\qm $\textup{PCA}_{1}$} and {\qm $\textup{PCA}_{2}$}, in log scale. The plots of the two classes were split to improve visibility that may otherwise be hindered by the overlap of the classes. {\qm The points are coloured by the probability $\mathbb{P}_{0}$ output by the first network in the ensemble.}}
    \label{fig:pcaprob}
\end{figure}

\begin{figure}[!]
    \includegraphics[width=1.01\columnwidth]{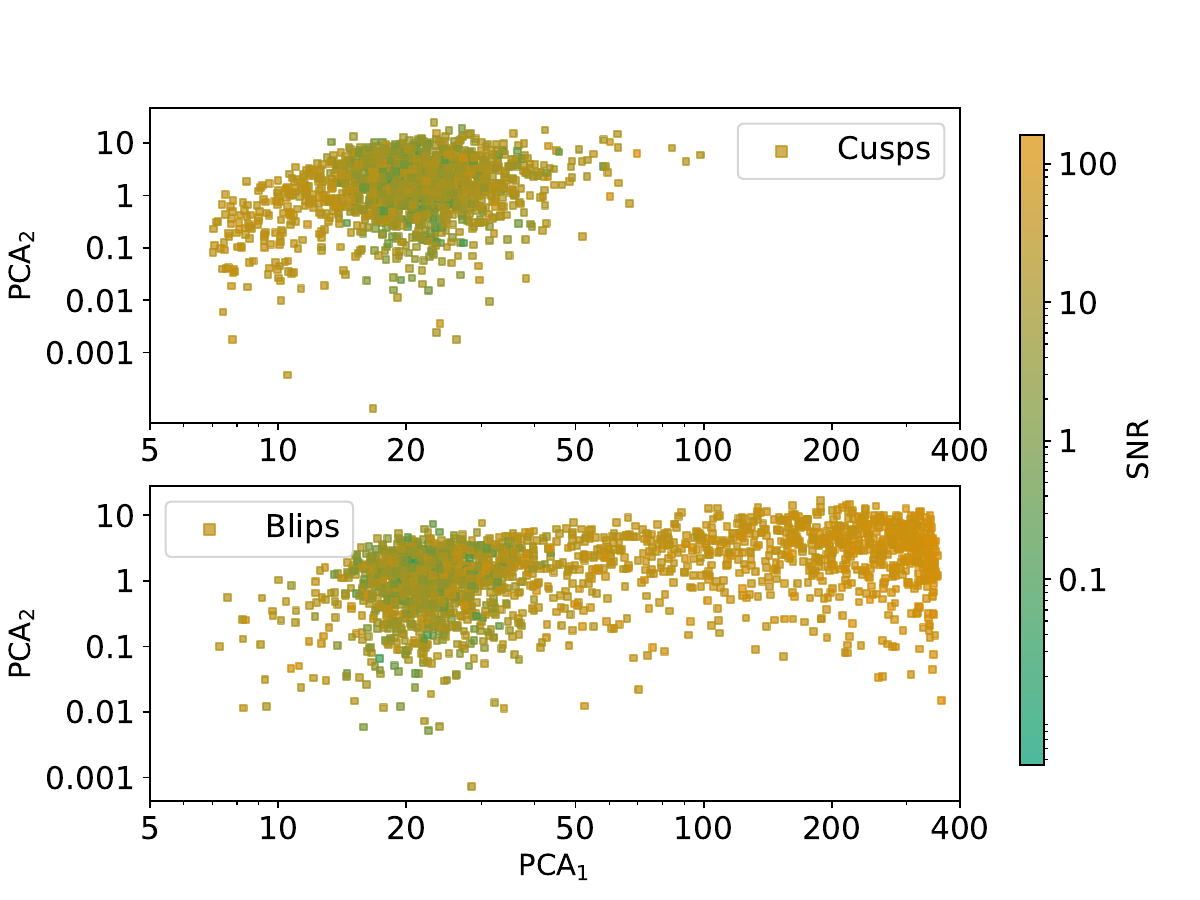}%
    \caption{A representation of a selection of activations from the dense layer in the space spanned by the first two principal components {\qm $\textup{PCA}_{1}$} and {\qm $\textup{PCA}_{2}$}, in log scale. The plots of the two classes were split to improve visibility that may otherwise be hindered by the overlap of the classes. The points are coloured by SNR.}
    \label{fig:pcasnr}
\end{figure}

For the first stream in the test set, the activations of the dense layer were projected onto the subspace spanned by the first two principal components. The results from this projection are shown in Fig. \ref{fig:pcaprob} and Fig. \ref{fig:pcasnr}, coloured by the probability $\mathbb{P}_{0}$ {\qm output by the first network in the ensemble} and the SNR, respectively. These figures are shown in log scale to improve the visual separation between the two classes. Both figures show a portion of the signal population being located in the top left of the plot, whereas a portion of the glitch population is located in the top right. Both classes overlap in the center and are therefore plotted separately to improve visibility. It is relevant to note that in the principal component space, glitches show a larger spread than the signals. This follows from their more varied possible morphologies.

It can be observed from Fig. \ref{fig:pcaprob} that the probability is related to the first principal component on the x-axis. Compared to Fig. \ref{fig:pcasnr}, the extremes of this same principal component show high values for the SNR. From this, it can be inferred that at least within this representation, the signals and glitches exhibiting the most separability are the ones that are loudest and therefore most obvious to the model. The first principal component can thus be interpreted as a measure of the example class. For the second principal component, there is no such apparent meaning.

\section{Conclusions}\label{sec:conclusions}

A deep-learning model that can distinguish between cosmic string cusp signals and blip glitches with significant accuracy was designed and analysed. Given that matched-filter searches for short-duration gravitational-wave signals are heavily hindered by short transient glitches such as blip glitches, the exploration of this task is important for both current and future searches. In this work, both populations were scaled to follow the same SNR distribution, meaning loudness was removed from the equation. With remarkable results for the accuracy ($79\%$) and true positive rate ($76\%$) in particular, it has been shown that deep learning is a viable candidate for use in cosmic string searches. Moreover, due to the classification speed of $10$ milliseconds per three data streams of $8$ seconds, the deep-learning model is fast enough to run as part of a real-time detection pipeline.

On a dataset consisting of injected signals and injected glitches, the deep-learning model was shown to outperform matched filtering at the task of distinguishing strains including signals from strains including glitches, winning mostly on the volume of false positives (as can be seen {\qm from the model's slow increase in true positive rate in} Fig. \ref{fig:roc}). This demonstrates that the deep-learning model is significantly better at rejecting glitches.

The behaviour of neural networks is notoriously difficult to understand, earning them the name of black-box models. As evidenced by their proven effectiveness, however, these black boxes hide valuable information. The hidden representations within the deep-learning model were interpreted through the application of three interpretability methods. The first of these is the method of waveform surgery, introduced in this paper, where parts of a waveform are removed to study the effects on the classification of such a procedure. The second method is a routine developed in this paper for the visualisation of convolutional filter activations for one-dimensional time series. The third is principal component analysis. These interpretations have resulted in several observations that may prove useful in future work. The glitch surgery procedures demonstrate the possibility of dividing waveforms into sections and show that models can be sensitive to changes within these sections. Surgery procedures can therefore be used to study the importance of distinct sections of waveforms to their classification. By considering a comparison between waveforms based on their sections, the complexity of signal discrimination can be reduced, therefore potentially reducing the difficulty of the task. Through the study of the convolutional filter activations, these latter values can be connected to the classes in the confusion matrix. Such studies may aid in making informed choices for convolutional filters, or more generally in neural network design. Lastly, principal component analysis applied to the throughput of the second to last dense layer of the network has enabled the study of class separability and the hidden procedure reducing the internal representation of the deep-learning model to output probabilities.

In the process of interpreting the deep-learning model, the morphological differences between cosmic string cusp signals and blip glitches were considered from the point of view of the model. Because the model was not allowed to rely on coincidence, it was fully dependent on these morphologies, yielding unique insight. At the same time, this serves as a proof of concept for high classification performance before coincidence is introduced to further improve a pipeline.

There are several directions open for continued work, such as the inclusion of other gravitational-wave generating mechanisms on cosmic strings. These mechanisms are comprised by kinks and kink-kink collisions, signals with different spectral indices from cusps that are now starting to be considered in cosmic string searches \cite{Abbott_2018, LVKO3CS}. Furthermore, there are indications the proposed deep-learning model offers additional room for improvement, for instance through an extended training phase. It is expected this adjustment will also serve to lower the false positive rate. In terms of analysis, the surgery process can be further refined, for instance by working at a resolution higher than three sections. This can be achieved by redefining the function of the standard deviation that marks the incisions.

Altogether, it is expected that the Einstein Telescope will bring a variety of new opportunities for the detection of cosmic strings and that deep learning will play a vital role in their analysis.

\section*{Acknowledgements}

The authors thank Adrian Helmling-Cornell and the anonymous referee for their helpful comments. With thanks to Artim Bassant for being open to string-theoretical discussion. Q.M and M.L. are supported by the research program of the Netherlands Organisation for Scientific Research (NWO). D.T. is supported by the Sherman Fairchild Postdoctoral Fellowship at Caltech. S.C. is supported by the National Science Foundation under Grant No. PHY-2309332. The authors are grateful for computational resources provided by the LIGO Laboratory and supported by the National Science Foundation Grants No. PHY-0757058 and No. PHY-0823459. This material is based upon work supported by NSF's LIGO Laboratory which is a major facility fully funded by the National Science Foundation.

\bibliography{references}

\onecolumngrid

\end{document}